\documentclass[10pt,journal,compsoc]{IEEEtran}
\usepackage{amsmath,amsfonts}
\usepackage{algorithmic}
\usepackage{array}
\usepackage{pifont}
\usepackage[caption=false,font=normalsize,labelfont=sf,textfont=sf]{subfig}
\usepackage{textcomp}
\usepackage[dvipsnames]{xcolor}
\usepackage{stfloats}
\usepackage{url}
\usepackage{hhline}
\usepackage{multirow}
\usepackage{makecell}
\usepackage{mathtools}
\usepackage{float}
\usepackage[hidelinks,colorlinks,allcolors=blue,urlcolor=black]{hyperref}
\usepackage{adjustbox}
\usepackage{colortbl}
\usepackage[numbers]{natbib}
\usepackage{tabularray}
\usepackage{verbatim}
\usepackage{vcell}
\usepackage{tikz}
\usepackage{cite}
\usepackage{balance}
\usepackage{orcidlink}
\usepackage{graphicx}
\usepackage{longtable}
\usepackage{comment}
\usepackage{amsmath,amsfonts}
\usepackage{algorithm}

\usepackage{stfloats}
\usepackage{url}
\usepackage{verbatim}
\usepackage{graphicx}
\usepackage{hhline}
\usepackage[numbers]{natbib}
\usepackage{authblk}  
\usepackage{ragged2e}

\ifCLASSINFOpdf
\else
\fi

\begin{document}



\title{False Alarms, Real Damage: Adversarial Attacks Using LLM-based Models on
Text-based Cyber Threat Intelligence Systems}

%
%
%
%

\author{Samaneh Shafee, Alysson Bessani, Pedro M. Ferreira}

\affil{
\texttt{sshafee@fc.ul.pt},
\texttt{anbessani@fc.ul.pt}, \texttt{pmferreira@ciencias.ulisboa.pt}\\
LASIGE, Faculty of Sciences, University of Lisbon, Lisbon, Portugal
}

\IEEEtitleabstractindextext{%

\begin{abstract}
\justifying
Cyber Threat Intelligence (CTI) has emerged as a vital complementary approach that operates in the early phases of the cyber threat lifecycle.
CTI involves collecting, processing, and analysing threat data to provide a more accurate and rapid understanding of cyber threats.
Due to the large volume of data, automation through Machine Learning (ML) and Natural Language Processing (NLP) models is essential for effective CTI extraction. 
These automated systems leverage Open Source Intelligence (OSINT) from sources like social networks, forums, and blogs to identify Indicators of Compromise (IoCs). 
Although prior research has focused on adversarial attacks on specific ML models, this study expands the scope by investigating vulnerabilities within various components of the entire CTI pipeline and their susceptibility to adversarial attacks.
It particularly focuses on a system-level analysis of how existing adversarial techniques interact with and across multiple stages of this pipeline, resulting in cascading attack effects.
These vulnerabilities arise because they ingest textual inputs from various open sources, including real and potentially fake content.
We analyse three types of attacks against CTI pipelines - evasion, flooding, and poisoning and assess their impact on the system’s information selection capabilities
Specifically, focusing on fake text generation, the work demonstrates how adversarial text generation techniques can create fake cybersecurity and cybersecurity-like text that misleads classifiers, degrades performance, and disrupts system functionality.
The focus is primarily on the evasion attack, as it precedes and enables flooding and poisoning attacks within the CTI pipeline.
Our findings reveal that the False Positive Rate (FPR) for the evasion attack reached 97\% for a specialized ML classifier model, indicating the model’s high vulnerability to adversarial samples.
Additionally, an FPR of 75\% is observed for ChatGPT-4o as a classifier, indicating its susceptibility to adversarial examples.
These results underscore the need for an additional verification component at the early stage of the CTI pipeline to detect and filter out misinformation before it spreads through the system.
\end{abstract}

}

\maketitle



\ifCLASSOPTIONcompsoc
\IEEEraisesectionheading{\section{Introduction}\label{sec:introduction}}
\else

\section{Introduction}
\label{introduction}
\fi
\IEEEPARstart{T}{he} 
increasing complexity of cyber threats has made it difficult for existing security measures to provide effective protection \citep{rahman2023attackers}.
Security systems such as firewalls and Intrusion Detection Systems (IDS) often cannot prevent breaches and fail to detect the spread of malicious activities.
An enhancement is to keep defense mechanisms up to date with appropriate configurations while enriching them with relevant and up-to-date threat information.
One promising approach to enhance defensive capabilities is Cyber Threat Intelligence (CTI), which provides insights into the tactics, techniques, and procedures (TTPs) used in modern cyberattacks, along with the corresponding detection and mitigation strategies.
CTI is an emerging field that analyses trends in cybercrime, hacktivism, and cyber spying by utilizing diverse sources, including open-source intelligence (OSINT), social media, and human intelligence \citep{samtani2017exploring}. 
An important challenge for CTI systems is the potential for attackers to inject fake intelligence across OSINT sources such as Twitter (currently named X\footnote{We use Twitter through the paper as our main dataset was published before the name change.}), Stack Overflow, dark web forums, and blogs, which can compromise the reliability of the intelligence gathered and pose challenges for organizations using such sources.
Although adversary-driven poisoning of CTI platforms (as in a confirmed incident) is rarely disclosed publicly, there are highly relevant cases of misinformation, fabricated cyber narratives, and deception that affected CTI platforms~\citep{Will2025Sans}.
Examples of the dissemination of fake CTI reports regarding ransomware, governmental organizations, and large companies have been reported~\citep{li2024web}.
For instance, threat actors and scammers on web forums have repeatedly claimed major data breaches or offered “fresh” stolen datasets that turn out to be fake or repackaged old leaks. 
These false breach narratives often get re-reported by media or aggregated into CTI feeds before verification, causing wasted defensive effort and misprioritization \citep{abrams_europcar_nodate}.

To address this, CTI pipelines have typically relied on trusted sources to reduce the risk of ingesting fake or manipulated information.
More broadly, CTI data collection in the literature has followed two main approaches: (1) keyword-based extraction \citep{shin2021twiti,bose2019novel} and (2) source-based curation from trusted entities \citep{alves2020follow,zhao2020timiner}. 
Keyword-based methods do not address the problem of misinformation, as they are vulnerable to ambiguous or misleading terms.
Relying on trusted sources also introduces limitations: their accuracy and availability may degrade over time, and maintaining an up-to-date trusted source list is difficult, especially when such sources become compromised or outdated.

This work explores the behavior of CTI pipelines when data is collected from diverse and potentially unverified sources.
This alternative approach avoids the limitations of relying solely on trusted sources and opens the possibility of exploring dark web forums. 
However, it becomes more vulnerable to misinformation and deliberate manipulation by adversaries.
The CTI extraction tools' failure to filter false or misleading information makes them vulnerable to adversarial exploitation. 
Attackers can manipulate these systems and compromise the integrity of the threat intelligence gathered.

In this paper, we design an integrated CTI pipeline capturing the essential stages found across existing systems.
This integrated pipeline serves as a unified reference framework for assessing the CTI pipeline against adversarial manipulation.
 
Despite their sophisticated workflows \citep{schlette2021comparative}, each pipeline stage remains susceptible to adversarial attacks, such as evasion, flooding, and poisoning. 
These attacks can severely impact the reliability of the pipeline, leading to incorrect predictions by ML models, monitoring and validation disruptions, and increased False Positive (FP) and False Negative (FN) outcomes.

While prior work has emphasized the benefits of automation in CTI extraction by ML models, the impact of adversarial manipulation on data sources has received limited attention.
Most existing CTI security research focuses on model-centric defences~\citep{Sarker2023Multi‐aspects}, emphasizing classifier robustness, entity extraction accuracy, or isolated defense mechanisms, and implicitly assumes that data sources are trustworthy \citep{Saeed2023A,Alaeifar2024Current,Nguyen2025Towards}. 
However, the systemic risks introduced by adversarial text generation at the data ingestion level and its compound effect on CTI pipelines remains insufficiently studied.
This work focuses on this overlooked intersection, where automation interacts directly with deception. 
It assesses the vulnerabilities of the CTI extraction pipeline, rather than introducing entirely new attack methodologies, aiming to demonstrate how adversaries can create a platform to deploy fake data and systematically deceive CTI models and systems. 
Besides the fake text generation methodology, the novelty of this study lies in a system-level examination of the cascading effects caused by existing adversarial text generation techniques across the CTI pipeline.
Understanding these deception mechanisms allows for evaluating the robustness of automated threat intelligence extraction processes more effectively.
In summary, the contributions of this work are as follows:
\begin{enumerate}
    \item {An integrated CTI extraction pipeline model is proposed that captures and generalizes the typical stages and components used across prior CTI pipelines.}
    \item {An attention-based fake text generation methodology using LLM chatbots and prompt optimisation.}
    \item {System-level analysis of structural vulnerabilities in CTI pipelines, focusing on how weaknesses arise from interactions between the interconnected stages.}
    \item{A comprehensive characterization of adversarial threats targeting automated CTI systems, including scenarios where attackers intentionally induce false positives by injecting misleading intelligence into data sources.}
\item {An evaluation of successive adversarial effects by applying evasion, flooding, and poisoning attacks and analyzing their combined impact on the pipeline performance.}
\end{enumerate}
Each contribution addresses a distinct limitation in the existing CTI security research.
Together, they provide a unified analysis of adversarial risks on a CTI pipeline model that are often examined separately.
By shifting the focus from individual models to interactions across the whole CTI pipeline, this study demonstrates how adversarial effects can accumulate over successive stages.
The fake text generation methodology describes realistic misinformation approaches that rely on generative chatbot LLM models rather than constrained perturbations.
Finally, the evaluation demonstrates that evasion attacks enable downstream flooding and poisoning, revealing system-level vulnerabilities that are not visible in prior work model-centric analysis.

The remainder of the paper is organized as follows. 
Section \ref{preliminary} introduces key definitions establishing the basis for the study.
Section \ref{cti-pipline} explains the five stages of the proposed CTI pipeline and Section \ref{Potential} presents a deep exploration of potential attacks on CTI pipelines.
Section \ref{Methodology} focuses on the progress of implementing evasion, flooding, and poisoning attacks.
The experimental results are presented in Section \ref{Experimental-Results}, along with our findings. 
Section \ref{Discussion} discusses the study's challenges, opportunities, and future work. 
Section \ref{Related-work} reviews related studies about adversarial text generation and adversarial attacks on ML models.
Finally, Section \ref{Conclusion} summarizes the key contributions and insights derived from this study.

\section {Preliminary definitions}
\label{preliminary}

This study focuses on a proposed CTI system that ingests textual data from OSINT to generate actionable alerts.
This section provides fundamental definitions related to the input textual data types and their predicted outputs by ML models, which are the core of CTI systems.

\noindent \textbf{{Real vs. Fake text.}}
It is important to distinguish between real and fake input text in CTI systems, but there are no universally standardized definitions.
Typically, text generated by a machine is labeled as fake, while human-written text is considered real \citep{crothers2023machine}.
However, this study adopts a task-specific definition:  
real text refers to genuine security-related information that reflects actual cyber threats or vulnerabilities; fake text refers to false, misleading, or manipulated cybersecurity information crafted to deceive either humans or automated systems. 
Such fakes may use cybersecurity-like terminology while not corresponding to real-world threats.
For example, a real security-related tweet is:  
\textit{"Vulnerability Details: CVE-2024-52046 (CVSS 10/10) Apache MINA Remote Code Execution (RCE) Vulnerability."}  
A fake version could be:  
\textit{"Exploit released for CVE-2024-52046 allowing full control over Apache MINA servers. No patch available yet—act now to secure your systems!"}. 
Although the fake version mimics the style and vocabulary of a real CTI statement, it conveys a false sense of urgency and misleading content.

\begin{figure}[ht]
    \centering
    \includegraphics[width=\columnwidth]{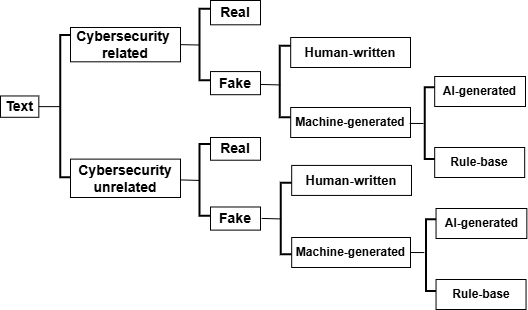}
    \caption{Taxonomy of input text in a CTI Pipeline.}
    \label{real-fake}
\end{figure}

Figure~\ref{real-fake} illustrates the taxonomy of the text input used in this study.  
First, the text is classified as related or unrelated to cybersecurity. 
Then, the taxonomy distinguishes fake texts based on whether it is a human-written or machine-generated text. 
Finally, fake machine-generated texts divides into two categorises including AI-generated or Rule-based.

\noindent \textbf{ {True vs. False Positives in CTI Classification.}}
To evaluate the effectiveness of classification models used in the CTI pipeline, we must distinguish between correct and incorrect prediction outcomes.
This study uses classic True Positive (TP) and False Positive (FP) definitions, focusing on binary classifier models that differentiate between security-related and unrelated texts. 
A TP reflects an accurate classification of a genuine CTI instance, while an FP often arises in adversarial scenarios where fake or unrelated content is intentionally crafted to resemble CTI text.
For example: \textit{"Vulnerability Details: CVE-2024-52046 (CVSS 10/10) Apache MINA Remote Code Execution (RCE) Vulnerability."} is a TP if correctly classified as CTI.
In contrast: \textit{"System Upgrade Details: CVE-2025-527656 (Performance Rating: 10/10) Apache MINA Remote Configuration Expansion (RCE) Module."} is an FP if the model misclassifies this fabricated security-irrelevant text as a real CTI alert due to the use of technical jargon and CVE-like formatting.
This distinction is critical for analysing model performance, particularly under adversarial testing scenarios that seek to exploit the classifier's reliance on superficial lexical cues.

\section{Text-based CTI pipeline}
\label{cti-pipline}
\begin{figure*}[]
    \centering
    \includegraphics[width=\linewidth]{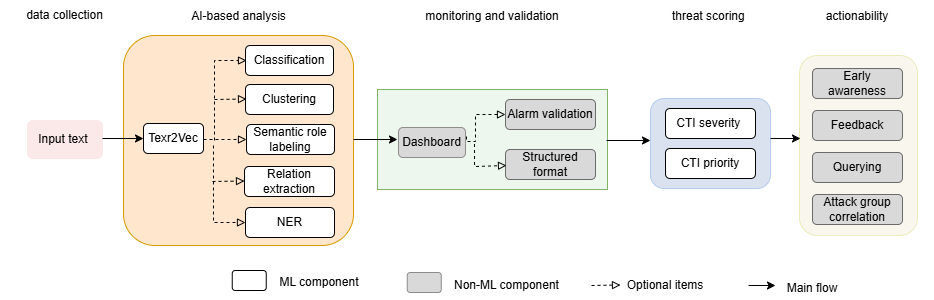}
    \caption{Proposed integrated CTI extraction pipeline.}
    \label{cti-view}
\end{figure*}

CTI extraction involves several key stages necessary to identify and extract relevant information, such as IoCs and TTPs.
Although many CTI pipelines have been proposed, they typically feature custom stages and components customized for specific organizational or research objectives.
An integrated text-based CTI extraction pipeline is proposed that unifies multiple extraction approaches into a structured framework.
As depicted in Figure \ref{cti-view}, it consists of five main stages: data collection, AI-based analysis, monitoring and validation, threat scoring, and actionability.
This design captures functionalities observed in prior CTI systems while consolidating them into a unified framework suitable for empirical evaluation under adversarial attack scenarios.
The following subsections describe each stage as adapted from prior works to form the integrated pipeline.
\subsection{Data collection}
At the data collection stage, raw textual inputs are gathered from various sources, including social media platforms, security blogs, vendor reports, CVE, and threat databases.
These inputs may contain early IoCs, emerging threats, or discussions on vulnerabilities.
The reliability and diversity of the collected data play a vital role in shaping the downstream analysis.
Some CTI systems restrict data collection to trusted sources to minimize FPs. 
In contrast, others broaden their scope to include informal or unverified channels, which may enhance coverage but could introduce greater data variability and require additional filtering.


\subsection{AI-based analysis}
Researchers have employed NLP techniques alongside ML models to extract CTI from textual data. 
Although extraction methods vary depending on the type of information, the overall CTI extraction process in the pipeline remains the same. 
This applies whether extracting IoCs, TTPs, or other relevant threat intelligence components.
Figure \ref{cti-view} provides an overview of techniques commonly used in CTI extraction pipelines, including classification, clustering, Named Entity Recognition (NER), semantic role labeling, and relation extraction \citep{shin2020new,alam2023looking,jo2022vulcan,satvat2021extractor,zhao2020timiner,rastogicyber,dionisio2020towards,alves2021processing,noor2019machine,long2019collecting}. 
These techniques enable models to identify patterns, extract key entities, and map relationships in text.  
Table~\ref{tab: cti-approach} lists specific examples applied in various studies that use diverse data sources and describes the contribution of the techniques to the purpose of the CTI tasks.
It is important to note that the implementation details and the sequence of techniques may differ across studies, depending on the specific goals and datasets used.

 \begin{table*}
\centering
\renewcommand{\arraystretch}{1.4}
\caption{Overview of AI-based methods for extracting CTI from textual data.}
\label{tab: cti-approach}
\resizebox{\textwidth}{!}{%
\begin{tabular}{llll} 
\hline
\rowcolor[rgb]{0.902,0.902,0.902} 
\textbf{Ref} & \textbf{Data sources} & \textbf{Extraction Techniques} & \textbf{Purposes} \\ \hline
\citep{noor2019machine} & Publicly available incident reports & Classification   & Automate cyber threat attribution \\ 
\hline
\citep{long2019collecting} & Cybersecurity reports & Classification, NER   & Automate IoC detection \\ 
\hline
\citep{shin2020new} & Twitter, CVE, Wikitext dataset & Classification  & Detect cybersecurity-related text \\ 
\hline
\citep{zhao2020cyber} & \makecell[l]{CVE and ExploitDB, Security \\Blogs,  Hacker Forum Posts} & Classification  & \makecell[l]{Vulnerability similarity analysis, \\ IoC recognition} \\ 
\hline
\citep{zhao2020timiner} &  \makecell[l]{Security blogs, vendor bulletins, \\and hacking forums} & Domain recognizer  & \makecell[l]{IoC recognition, \\ Threat severity} \\ 
\hline
\citep{dionisio2020towards} & Twitter & Classification, NER   & \makecell[l]{Detect cybersecurity-related text \\ and extract cyber threat entities} \\ 
\hline
\citep{alves2021processing} & Twitter & Classification \& Clustering   & Detect aggregate cybersecurity-related text \\ 
\hline
\citep{satvat2021extractor} & CTI reports & Semantic role labeling  & \makecell[l]{Extraction of attack behavior, \\ Threat hunting} \\ 
\hline
\citep{jo2022vulcan} & Publicly Available Sources & NER, relation extraction & \makecell[l]{Knowledge graph construction from \\ extracted entities and their relationships \\ in a graph database} \\ 
\hline
\citep{alam2023looking} & CTI Reports & TTPClassifier, NER   & \makecell[l]{Attack Pattern Predictions, \\ Transform unstructured CTI information \\ into a structured knowledge graph} \\ 
\hline
\citep{rastogicyber} & Threat analysis reports  &  \makecell[l]{Classification and Sequence\\ Tagging, NER, relation extraction}  & \makecell[l]{Detect semantically similar \\ attack patterns} \\ 
\hline

\citep{cui2023atdg} & Dataset of ref \citep{dionisio2020towards} & Classification, NER & \makecell[l]{Detect cybersecurity-related text \\ and extract cyber threat entities}  \\ 
\hline

\end{tabular}%
}
\end{table*}
\subsection{Monitoring and validation}
The monitoring stage comprises key components such as the dashboard, alarm validation, and a structured format. 
The dashboard provides a visual overview of recent events, which helps security operations center (SOC) users maintain situational awareness and respond to threats, as supported by research \citep{alves2021processing} and tools such as \citep{emk_cti_nodate}. 
These implementations underscore the importance of dashboards in centralizing information and supporting decision-making in cybersecurity operations.
Alarm validation is a valuable task performed by a skilled human analyst who reviews and verifies alerts generated earlier in the pipeline \citep{Afzaliseresht2020From}.
The validated alerts are then organized and shared in structured formats such as STIX or TAXII \citep{rahman2022threat, Briliyant2021Towards}.
\subsection{Threat scoring}
This stage evaluates the severity and priority of detected threats \citep{alam2024ctibench}.
Severity reflects the potential impact based on factors such as exploitability and system exposure, while priority defines the response order based on severity and the likelihood of exploitation \citep{kerkdijk2021evidence}.
Priority helps allocate resources effectively and ensures a timely response \citep{jalalvand2024alert}.
Severity is typically labeled as Critical, High, Medium, or Low, whereas priority includes Urgent, ASAP, Within 24 hours, and Low Priority levels \citep{Vielberth2020Security}.
This distinction improves decision-making and response efficiency in SOCs.
\subsection{Actionability}
This stage of the CTI pipeline builds on foundational components to enable organizations to anticipate, analyse, and act on threats. 
Key functions include:

\noindent \textbf{Early awareness.} 
Extracting OSINT CTI from online platforms enables security professionals to identify potential threats before they fully emerge. 
Alves et al. \citep{alves2020follow} demonstrated that security-relevant tweets can be detected up to 148 days before NVD vulnerability disclosure, allowing proactive defense and risk mitigation.
 
\noindent \textbf{Querying CTI.} 
Structured CTI allows efficient querying of relevant information \citep{piplai2020creating, neil2018mining}.
For example, CTI extracted from hacker communities can be organized in a searchable portal where users can filter data by time or resource type to identify specific patterns or events \citep{samtani2016azsecure, samtani2017exploring}. 
Such systems improve the accessibility and usability of CTI and enable security teams to make more effective, informed decisions.
Structured querying of security knowledge graphs also identifies vulnerabilities in software libraries before deployment.
 
\noindent \textbf{Feedback.} 
Some CTI systems use feedback loops in which SOC teams evaluate the quality of extracted data, filtering out inaccuracies and improving future extractions by retraining models against quality standards \citep{zvelo_cyber_2020}.
 
\noindent \textbf{Attack group correlation.} 
The extracted CTI helps link patterns between incidents with known threat actors and provides information on their tactics and priorities.
This correlation helps identify threats and allows defense contractors to prioritize responses based on the attacker's behavior and capabilities~\citep{li2024advanced}.

\section{Attacks on CTI pipeline}
\label{Potential}
Each component of the CTI pipeline can become a potential target for adversaries aiming to compromise it.
Therefore, we analyse attacks targeting automated text-based CTI pipelines, including their capabilities, knowledge, and goals.

\subsection{Overview of attacks}
This study examines three possible attacks to illustrate key threats.
The objective is to provide a concise reference that explains the mechanisms and implications of each attack and demonstrates their impact on various system components, including ML models, the dashboard, and alarm validation.

\noindent \textbf{{Evasion attack.}}
Evasion attacks occur during the test step, with trained models targeted to be misled.
Once the extractor pipeline is trained, adversaries can generate Fake true Negative (FaN) input texts to fool the model into misclassifying them as positives.
This manipulation leads to increased FP and FN texts appearing on the dashboard, confusing SOC professionals and diverting attention from real threats \citep{pitropakis2019taxonomy}.
In the test scenario, this attack focuses on generating FaN inputs that resemble real cybersecurity texts but are intentionally designed to cause misclassifications. 
For example, for a binary classifier that distinguishes between security- and non-security-related texts \citep{dionisio2020towards}, an adversary can lead the model to misclassify non-security texts as security-related, possibly disrupting decision-making.
This attack is expressed as,
\begin{equation*}
\arg\max_{\hat{x}}\left\{\mathrm{Loss}\left(f\left(\hat{x}\right), y\right)\right\}\,,
\end{equation*}
where $f(\hat{x})$ indicates a binary classifier that predicts whether an input text is security-related or non-security-related, $\hat{x}$ represents the adversarially modified input text, and $y \in \{0,1\}$ refers to the ground-truth class label (0: non-security-related, 1: security-related).
In this work, the loss function is the binary cross-entropy.
The adversary seeks to modify the original input $x$ into $\hat{x}$ to maximize the classification loss, thereby increasing the likelihood of misclassification at inference time.
Evasion attacks can be categorized into two main approaches: maximum-confidence and minimum-distance \citep{Li2020Adversarial}. 
Maximum-confidence attacks aim to create adversarial examples that are misclassified with high confidence, but often involve substantial changes to the input that decrease its similarity. 
In contrast, minimum-distance attacks focus on minor and subtle modifications to maintain high similarity and make detection more difficult.

The minimum distance approach is adopted, since preserving the similarity of adversarial texts is crucial to bypass classifiers and aligns with the need for realistic adversarial examples in cybersecurity.
Evasion attacks are a foundation for other attacks.
A successful evasion attack increases the risk of poisoning attacks, where FPs enter the training set and corrupt future model retraining.
Additionally, increased FPs can flood analyst workflows, leading to delayed responses to real security threats.

\noindent \textbf{Flooding attack.}
To implement our flooding attack, we need to generate manipulated real cybersecurity texts known as Fake true Positives (FaP).
Flooding attacks \citep{7456891} overwhelm the CTI extraction pipeline by injecting a high volume of deceptive FaN and FaP texts into the system dashboard, making it difficult for security analysts to distinguish between real and misleading alerts.
Additionally, attackers may use this technique to conceal their malicious activities within the flood of data to evade detection. 
This strategy can be considered a form of denial-of-service (DoS) attack, where the system becomes overwhelmed with excessive traffic, preventing it from processing legitimate requests. 
As a result, real positives become obscured, diminishing the system's ability to handle real threats.
This overload weakens the reliability of the threat intelligence pipeline and increases ambiguity in threat detection, making it harder for analysts to respond to genuine risks. 
Beyond disrupting decision-making, flooding attacks waste the analyst’s effort, leading to system failure and degrading the overall security response efficiency.
Repeatedly encountering false alerts results in a loss of trust in the model’s predictions, reducing confidence in automated classification outcomes. 
Furthermore, overwhelming SOC analysts strains security operations and effectively impairs their ability to manage real threats. 
The excessive volume of injected data also increases costs owing to log storage, requiring more computational resources to process and archive irrelevant information.
This not only affects infrastructure overhead but also directly impacts the analyser’s ability to process and interpret incoming data accurately.
When encountering fake inputs, the analyser has two possible reactions: discard them, wasting computational resources and analyst time; or misclassify them as genuine, leading to a poisoning attack.

\noindent \textbf{Poisoning attack.}
Poisoning attacks \citep{cina2024machine} manipulate training data to degrade CTI extraction model accuracy and are classified into integrity and availability attacks. 
Integrity attacks alter specific data points to cause targeted misclassification. 
Availability attacks modify large portions of the training set and reduce overall model reliability \citep{tian2022comprehensive}. 
These attacks primarily target the system's test step and training set components. 
Evasion attacks can facilitate poisoning by contaminating the training data and allowing access to undetected FaN or FaP texts. 
If these are misclassified as TPs or TNs, respectively, they corrupt the model's learning process and degrade model performance. 
The consequences can be significant, generating false security alerts, spreading misinformation across interconnected systems and professionals, and causing analysts to respond to non-existent threats while overlooking real ones.

\subsection{Attacker conceptual model}
Understanding the knowledge, capabilities, and objectives of the attacker is essential for modeling realistic threat scenarios, including how these factors appear in evasion, flooding, and poisoning attacks.
These factors influence how and where the adversary may attempt to compromise the CTI pipeline, as different attacker profiles adopt different strategies.

\noindent\textbf{Attacker’s capability.}
We explain the attacker capability requirements, classifying attacks by increasing thresholds of required resources, persistence, and influence over the CTI pipeline.
Specifically, we distinguish three attacker capability levels: low-capability, mid-capability, and high-capability.

In the evasion attack, the attacker generates adversarial inputs with minimal detectable changes to exploit model vulnerabilities at inference time.
This low-capability attack can be conducted at inference time using publicly accessible tools, does not require access to model internals, and requires the least attacker capability.

In the flooding attack, the attacker injects coherent false data streams to overwhelm the system, disrupting data integrity and operational efficiency.
Such mid-capability attacks require moderate computational resources and sustained activity to continuously flood the data ingestion process.
Additionally, in a poisoning attack, the attacker generates FaN texts that closely resemble real security-related texts, deceive both the ML models and analysts, and ultimately corrupt the training dataset.
This is a high-capability attack because it requires persistence over time, computationally optimised text generation, and the ability to generate diversified FaN texts that will pass the human analysts, reach the training data, and influence the model decision boundary.

\noindent\textbf{Attacker’s knowledge.}
The attacker's knowledge is formalized by a tuple $\kappa \in K$, where $K$ represents the abstract knowledge space encompassing all knowledge dimensions an attacker might have about the target. 
Specifically, $\kappa = \left(D, X, f, L, w\right)$, where $D$ denotes the training data, $X$ represents the feature set of the victim model, $f$ refers to the learning algorithm, $L$ is the training objective function, and $w$ corresponds to the trained model's parameters.
The attacker cannot directly access $D$ but may approximate it using publicly available OSINT datasets, such as cybersecurity repositories (e.g., CVE) that could overlap with $D$. 
Using these resources, the attacker constructs a substitute model to mimic the target system's behavior.  
While features $X$, $f$, and $L$ remain unknown, the attacker can infer likely ones based on common cybersecurity and ML practices. 
Finally, the trained model's parameters $w$ are entirely inaccessible. 
In this black-box attack scenario, the attacker's knowledge $\kappa$ is a constrained instance of $K$, reflecting the practical limitations attackers face.
Black-box attacks are more challenging than white-box or gray-box attacks because they rely only on input-output interactions without assuming any knowledge of the model \citep{bhambri2019survey}. 

\noindent\textbf{Attacker’s goal.}
In the evasion attack, the attacker aims to increase the percentage of FP, misleading the analyser and compromising the system's reliability.
Misclassification disrupts decision-making, increases errors, and risks of overlooking real threats hidden among FPs.
The flooding attack overwhelms dashboards with excessive FaN and FaP texts, shifting the focus from actual threats, causing system slowdowns, crashes, analyst fatigue, or infrastructure strain, ultimately reducing security effectiveness and costs.
In a poisoning attack, the attacker aims to corrupt the learning process of the model by generating deceptive texts that are mistakenly included in the training dataset.

\section{Methodology}
\label{Methodology}
This section outlines the methods for simulating evasion, flooding, and poisoning attacks in the CTI extraction pipeline, considering the data set, target models, text generation, and attack procedures. 

\subsection{Dataset}
A publicly available Twitter dataset with 31281 tweets \citep{dionisio2019cyberthreat} was the basis for generating FaN and FaP texts
,of which 11074 (35.4\%) are security-related and the remaining 20207 (64.6\%) are non-security tweets.
Although the dataset was originally collected across three time periods (2016, 2017, and 2018), it was later annotated for NER and republished by Dionisio et al. \citep{dionisio2019cyberthreat}.
This dataset offers unique advantages that make it suitable for this study.
It was employed to train the target model \citep{dionisio2020towards}, ensuring experimental consistency and compatibility with the attack scenario. 
It includes human-annotated named entities extracted from the cybersecurity-related tweets, essential for generating realistic FaP texts in the flooding attack setup.
To the best of our knowledge, no other publicly available CTI corpus simultaneously provides binary relevance labels and manually annotated named entities, both of which are required for designing our flooding and poisoning attacks.

The dataset includes a classification label ("relevant class") that indicates whether a tweet is security-related (1) or not (0). 
Although the tweet sizes of the dataset are limited to 256 characters, the FaP and FaN texts generated for evasion, flooding, and poisoning attacks vary in length. 
The clean\_tweet feature in the dataset, which contains refined tweet text, was used to create adversarial examples.
In addition to the Twitter dataset, we evaluated our approach using the AnnoCTR~\citep{langeannoctr} dataset of 400 CTI reports, which consists entirely of cybersecurity-related content collected between 2020 and 2024. 
The longer text lengths in AnnoCTR enabled the generation of more elaborate and detailed FaN texts.
From the dataset, which contains cyber threat reports segmented into individual text fragments, we sampled 2000 segments. 
Of these, for comparison purposes, 1500 were selected randomly to approximate the length distribution of Twitter posts (100-256 characters), while the remaining 500 consist of longer segments ranging from 256 to 1000 characters.

\subsection{Target models}
Two models are targeted in the experiments. 
The first is a state-of-the-art binary classifier developed by Dionisio et al. \citep{dionisio2020towards} based on Convolutional Neural Networks (CNNs), employed within CTI extraction pipelines to recognize whether a tweet is related to cybersecurity or not. 
Binary labels and classifiers are fundamental requirements for the evasion attack considered in this study, as the generation of FaN texts explicitly relies on a binary decision boundary between security-related and non-security-related inputs.
A successful evasion attack is defined as forcing a non-security text to cross this boundary and be misclassified as security-related, which cannot be unambiguously defined in multi-class or open-set settings.

We target the state-of-the-art binary classifier proposed by Dionisio et al. \citep{dionisio2020towards}, which is based on Convolutional Neural Networks (CNNs) and used in CTI extraction pipelines to distinguish cybersecurity-related tweets from others. 
This choice is motivated by its recognised effectiveness \citep{altalhi2021survey} and training on real-world tweets, ensuring a realistic adversarial target. 
Moreover, the binary nature is essential for reliably defining evasion success in FaN text generation, because multi-class or open-set settings lack a clear decision boundary.
The second target model is Chat( GPT-4o and GPT-5-mini), selected for exceptional performance in classifying security-related and non-security texts \citep{shafee2024evaluation}.
Unlike traditional classifiers, ChatGPT integrates contextual understanding with classification tasks to effectively detect subtle cyber threat patterns. 
Moreover, the model's widespread adoption and versatility make it a practical benchmark for testing adversarial robustness in real-world scenarios. 
By targeting ChatGPT, we aim to evaluate the resilience of cutting-edge language models against evasion attacks.
The evasion attack targets both models, while the flooding and poisoning attacks are specific to the specialized classifier \citep{dionisio2020towards}.
\subsection{Adversarial text generation}
\label{fake-text}
Adversarially generated FaN texts are crucial for all the attacks considered in this study. 
While previous research has explored adversarial text generation in non-security domains \citep{ranade2021generating,huynh2021argh,zellers2019defending,ren2020generating}, these approaches often lack public implementations or detailed documentation. 
Furthermore, while some of these methods are complex, the advent of large language models (LLMs) has simplified the process.
This study demonstrates that even publicly accessible LLMS, such as GPT, can be effectively prompted without fine-tuning, generating adversarial inputs that closely resemble cybersecurity-related texts.
They can mimic the features and structures of real text, and their advanced ability to follow instructions allows for a highly controlled and flexible generation process.
This observation reveals a major vulnerability in CTI pipelines, as they can be manipulated using generic tools without the need for sophisticated, domain-tuned generation frameworks.
To better understand how this vulnerability arises, this work investigates how LLMs can generate adversarial texts that closely mimic the language and structure of cybersecurity content.
Open-source LLMs like LLaMA 3, including the 8 billion (8B) and 70 billion (70B) parameter versions \citep{dubey2024llama} released in April 2024, were assessed for their potential in adversarial text generation.
However, they failed to meet quality requirements, lacking diversity and producing unrealistic adversarial samples.

\begin{figure*}
    \centering
    \includegraphics[width=\textwidth]{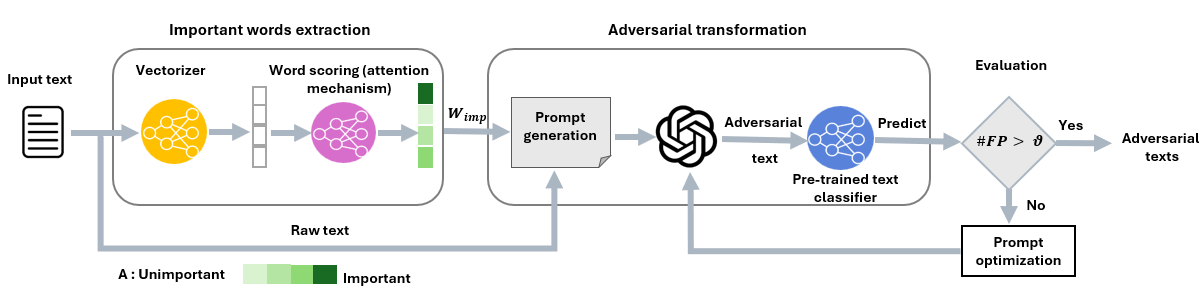}
    \caption{Illustration of generating adversarial text using the attention mechanism and ChatGPT-4o. 
    The generated adversarial text is input to a pre-trained binary classifier to mislead its predictions.
    A: The color intensity indicates the importance of words, ranging from less important (light) to highly important (dark). 
    The top three most important tokens in each tweet were incorporated into the prompt.}
    \label{general-view}
\end{figure*}

ChatGPT-4o was leveraged to generate texts that mimic the structure and terminology of cybersecurity texts while remaining unrelated to real cybersecurity.
All adversarial texts were generated using ChatGPT-4o and ChatGPT-5-mini, each with a fixed prompt that preserves the lexical appearance and structural patterns of cybersecurity texts while explicitly removing security-related semantics. 
To ensure reproducibility, we used fixed generation parameters with a temperature of 0.7 and a maximum length limited to 100 tokens.
While LLM-based text generation has limitations, such as potential biases, repetition, and inaccuracies \citep{chung2023increasing}, the method proposed in this study exceeded the expectations. 
To generate effective adversarial texts with ChatGPT-4o, prompts were designed using real tweets.
Prompt optimization was performed iteratively.
The initial prompt simply instructed the model to rewrite a cybersecurity-related text into a non-security text while preserving its surface structure and excluding security-related semantics.
However, early outputs often leaked security semantics or appeared linguistically generic.
Over five refinement iterations, we progressively introduced explicit constraints to:
(i) forbid any security-related content;
(ii) enforce transformation of key terms into visually similar but non-security tokens;
(iii) preserve the original sentence structure;
(iv) emphasize stylistic resemblance to CTI messages; and
(v) encourage lexical diversity during generation.
The final prompt, shown in Figure~\ref{prompt} in \ref{AppendixA}, reflects these refinements and produces texts that are non-security in content while remaining deceptively similar to cybersecurity messages in their surface structure, thereby reducing semantic leakage and improving stylistic deception.
An attention-based mechanism is proposed to extract the contextually key tokens from the real cybersecurity tweets.
In the following, we first define the problem of generating adversarial text and then explain the generation process and its components as shown in Figure \ref{general-view}.

\noindent\textbf{Problem Definition.}
The problem can be stated as: \textit{How can we generate FaN texts by modifying input text while preserving its cybersecurity-like semantics?}
Let $x=\left\{t_1, t_2, \dots, t_n\right\}$ represent a text sample consisting of $n$ tokens, where a token can be a word, subword, or special character, depending on the tokenizer's segmentation process. 
The sample $x$ belongs to the sample space $X$ and the target classification model \( F: X \to Y \) assigns \( x \) to a class \( y \in Y \), i.e., \( F(x) = y \). 

To generate FaN texts, key tokens \( t_{\text{imp}} \subset x \) are identified using an attention-based mechanism that quantifies the importance of each token based on its contribution to the model's decision.  
Unlike conventional adversarial attacks that directly perturb important tokens, the proposed approach preserves \( t_{\text{imp}} \), modifying instead surrounding tokens to shift the overall meaning while maintaining a cybersecurity resemblance.
The transformation process modifies tokens around \( t_{\text{imp}} \), ensuring \( t_{\text{imp}} \) remains unchanged to retain the structure of a cybersecurity-related text without conveying real security text.

\noindent\textbf{Important token extraction.}
Using a pre-trained SecureBERT model \citep{aghaei2022securebert}, the input texts are tokenized and embedded in vector representations.
SecureBERT is a specialized language model trained on cybersecurity corpora, enhancing the effectiveness in understanding and distinguishing security-related text \citep{levi2024cyberpal}.
The embeddings are processed by a Long Short-Term Memory (LSTM) layer to capture sequential relationships. 
The LSTM is a recurrent neural network designed to model temporal sequences and capture long-term dependencies, effectively addressing issues like the vanishing gradient problem \citep{graves2012long}. 
Subsequently, an attention layer assigns weights \( \alpha_i \) to each token \( t_i \), reflecting its contextual importance:
\[
\alpha_i = \frac{\exp(e_i)}{\sum_{j=1}^n \exp(e_j)}, \quad e_i = \text{tanh}(h_i W_a + b_a),
\]
where \( e_i \) is the intermediate relevance score of \( t_i \) in the given context, \( h_i \) is the hidden state of the LSTM, and \( W_a \) and \( b_a \) are the trainable parameters of the attention layer. 
The top-\( k \) tokens with the highest attention scores \( \alpha_i \) contribute the most to the model decision and are selected as \( t_{\text{imp}} \), the most influential tokens in the text.

\noindent\textbf{Adversarial transformation.}
First, based on empirical analysis to select the threshold, the three most important tokens,  \( t_{\text{imp}}=\left\{t_{\text{imp1}}, t_{\text{imp2}}, t_{\text{imp3}}\right\} \), are identified.
The threshold aims to balance the need for sufficient perturbation to influence the model's output while preserving the original semantic structure of the text.
Formally, given an input sequence of tokens,
\[
\begin{aligned}
x = \{ & t_1, t_2, \dots, t_{k-1}, t_{\text{imp1}}, t_{k+1}, \dots, t_{l-1},\\
       & t_{\text{imp2}}, t_{l+1}, \dots, t_{m-1}, t_{\text{imp3}}, t_{m+1}, \dots, t_n \}\,,
\end{aligned}
\]
a transformed text is generated,
\[
\begin{aligned}
\hat{x} = \{ & t_1^*, t_2, \dots, t_{k-1}^*, t_{\text{imp1}}, t_{k+1}, \dots, t_{l-1}^*, \\
             & t_{\text{imp2}}, t_{l+1}^*, \dots, t_{m-1}, t_{\text{imp3}}, t_{m+1}^*, \dots, t_n \}\,,
\end{aligned}
\]
where \( t_{\text{imp1}}, t_{\text{imp2}}, t_{\text{imp3}} \) remain unchanged to maintain key cybersecurity indicators.
Instead of modifying important tokens, ChatGPT-4o replaces some surrounding tokens, denoted by \( t^*_i \), with contextually similar but modified terms that eliminate security-related content while preserving the overall sentence structure.
The positions of \( t^* \) are not the same for all sentences. 
They are dynamically determined based on the text structure and context. 
To achieve this, ChatGPT-4o selects replacements from domains unrelated to security that are suggested in the prompt.
The modified text must meet two primary objectives:  
\textit{(1) Preservation of the textual structure and semantic similarity}, so the generated text closely resembles a cybersecurity-related text structurally and semantically;
\textit{(2) Changing the security-related nature of the text} to shift the content away from cybersecurity topics.

Figure \ref{general-view} illustrates the FaN text generation methodology\footnote{Code will be available on \url{https://github.com/samanehshf/Fake-CTIs}}. 
The attention mechanism identifies the top three tokens. 
In our experiments, the number of key tokens extracted using the attention mechanism was fixed to three.
This choice strikes a balance between preserving sufficient cybersecurity-related lexical cues and minimizing excessive semantic leakage from the original content.
Using fewer tokens (e.g., two) often resulted in adversarial texts that lacked sufficient stylistic resemblance to cybersecurity texts, whereas using more tokens (e.g., five) increased the risk of retaining security-specific semantics.
To validate this design choice, we conducted a series of smaller-scale experiments in which we varied the number of key tokens and evaluated their impact on prompt effectiveness. 
Across all settings, using three tokens led to a better deceptive effect and achieved the highest evasion performance.
Then ChatGPT-4o modifies the surrounding top three tokens as suggested in the prompt to create adversarial texts that still appear cybersecurity-related. 
The prompt follows an iterative refinement, where each cycle evaluates the generated FaNs based on their ability to deceive the pre-trained text classifier used in our pipeline.
The prompt is considered adequate if the classifier produces a number of FPs ($\#FP$) exceeding a predefined threshold $\vartheta$.
Otherwise, another refinement iteration is executed.
The finalized prompt was effective enough to surpass the $\vartheta$ setting of 80\% of the input texts.
It was then used to generate the adversarial FaN texts, forming the basis for subsequent evaluation.

\noindent\textbf{Human evaluation.} 
A human evaluation step was included after the final prompt optimization iteration to further validate the effectiveness of the generated FaN texts.
To filter out low-quality fakes, the data were distributed to three analysts who were informed that the texts were fake and tasked with identifying those that resembled cybersecurity texts.
The analysts had prior experience in cybersecurity-related research and threat intelligence analysis, and were asked to base their judgments solely on linguistic plausibility, domain-specific terminology, and overall resemblance to real-world CTI texts.
These three dimensions operationalize the notion of “resemblance” and provide a consistent basis for human judgment.
Analysts were clearly informed that the texts were fake to prevent external fact-checking or online verification, thereby enforcing a conservative evaluation setting in which texts had to appear realistic even under careful review.
This result was crucial to achieving a final prompt that surpassed the $\vartheta$ threshold.
The same analysts were later involved in downstream validation stages of the CTI pipeline, reflecting a realistic analyst-in-the-loop workflow and avoiding redundant human evaluations.
The importance of the iterative prompt refinement with human validation is demonstrated by the difference between the initial and final FPR presented in Section \ref{Experimental-Results}.


\subsection{Evasion attack}
\label{implement-Evasion-attack}
The evasion attack was conducted by passing the adversarially generated FaN texts through the target classifiers.
A successful attack occurs when a FaN is incorrectly classified as positive, resulting in an FP.
Otherwise, it is considered a TN. 
The effectiveness of the attack is quantified using the FPR, allowing an assessment of the vulnerability of the classifier to adversarially generated inputs at inference time.
In contrast to classical adversarial NLP attacks that rely on constrained token-level perturbations or explicit optimization of a loss function, the evasion attack considered in this work is generation-based.
It models a realistic CTI threat scenario in which adversaries inject fully fake cybersecurity-like texts into OSINT sources, rather than minimally modifying existing samples.
This design choice reflects real CTI pipelines, where attackers freely generate deceptive content at scale using public LLMs without edit budgets or semantic preservation constraints.
\subsection{Flooding attack}
The flooding attack is executed using a combination of FaN texts, paraphrased FaP, and rule-based generated FaP texts, to study the efficacy of different generation methods and text types.

\noindent \textbf{Paraphrasing TP texts.}
Multiple variations of a single real TP text are generated through paraphrasing.
These texts serve as an effective means of flooding the model with slightly different yet semantically similar inputs.
To achieve this, GPT-3.5-Turbo API was employed as a paraphrasing tool, applying it to the clean\_tweet feature of the dataset.
By generating ten paraphrases for each tweet from the positive class (1), 11074 FaP input texts were created.

\noindent \textbf{Rule-based fake texts.}
The generation of FaP texts using a rule-based approach involves replacing words based on their named entity type to preserve contextual consistency.
The tweets include entity features such as organization names, product names, vulnerabilities, and version information.
Among the 11074 positive tweets, 4552 include organization names, 10218 mention product names, 5137 contain vulnerabilities, and 3640 refer to version information.
To ensure coherence, the replacement words for each entity type are sourced from other occurrences of the same type within the dataset. 
This guarantees that substitutions remain contextually valid and do not introduce semantic inconsistencies. 
Entities within each entity type are categorized into predefined semantic groups, ensuring that replacements occur only within the same conceptual category.
For example, we categorized vulnerability entities into attack types, attributes, and execution methods. 
Within the attack type category, terms such as ransomware, trojan, and malware are substituted only with other attack types. 
This structured approach ensures the generated FaP texts maintain a realistic cybersecurity-related appearance. 

\begin{figure}[ht]
    \centering
    \includegraphics[width=\columnwidth]{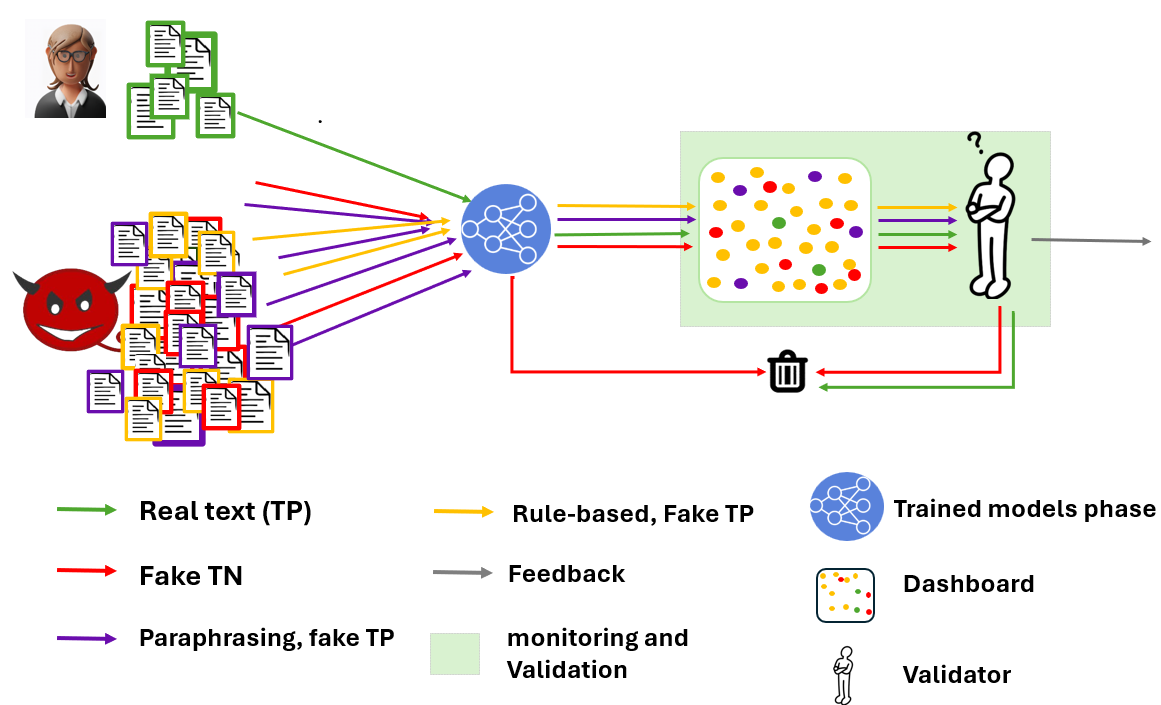}
    \caption{Flooding attack workflow.}
    \label{fig:flooding_attack}
\end{figure}

\noindent \textbf{Flooding attack mechanism.}
The mechanism of a flooding attack on a CTI extraction system is illustrated in Figure \ref{fig:flooding_attack}, showing the workflow for injecting FaN and FaP texts.
The honest security hacker on the left generates real positive threat intelligence texts (green arrows).
Conversely, an adversary generates and injects a large volume of fake texts of different types, designed to bypass the trained model.
The trained model filters some of the inputs but fails to reject all malicious inputs, leading to a significant number of FaP (green, yellow) and FaN texts (red) passing through and appearing on the dashboard, along with the real TP text (green).
The overwhelming volume of excessive and conflicting alerts confuses analysts, making it challenging to differentiate between real and deceptive inputs and leading to unnecessary investigation time and errors, such as discarding real threats (green) or leaving fake alerts in the system.

\subsection{Poisoning attack}
In the proposed CTI pipeline (Figure~\ref{cti-view}), a poisoning attack can occur through feedback in the actionability stage.
Alerts confirmed by analysts are incorporated into the dataset used to retrain the model, introducing a vulnerability that adversaries can exploit.
This poisoning attack gradually corrupts the retraining process, building upon successful evasion or flooding attacks described in previous subsections.

To demonstrate the impact of the poisoning attack, 9402 FaN texts that were misclassified as positive (FP) by the model \citep{dionisio2020towards} and human analysts in the experimental evaluation of the evasion attack were gradually injected into the model training dataset over several retraining rounds.
In each round, a random number of FaNs were uniformly sampled and removed from the 9402 set, and injected into the training data to simulate realistic and unpredictable successive poisoning conditions.
As a result, the classifier gradually begins to learn incorrectly as the cumulative number of FaN entries grows, affecting its classification decision boundaries and performance.

Although not demonstrated in this work, poisoning can also occur through flooding attacks, from which also FaP texts can reach the training set.
In this case, not only FPs affect model classification performance, but also FaPs can introduce model bias and data imbalance.

\subsection{Evaluation methodology}
The attacks studied are characterized in Table~\ref{text-classification} in terms of the adversarial text input type, purpose, and the corresponding classifier responses, explaining how the different attacks exploit the model's vulnerabilities and the evaluation metrics used in the experiments.
In evasion and flooding attacks, the adversary generates FaN texts to cause the model to misclassify them as FP samples, thus increasing the FPR.
In this context, the FPR directly corresponds to the attack success rate (ASR), as a successful attack is defined by the misclassification of adversarial inputs as security-related.
In flooding attacks, the adversary further intensifies the disruption by generating FaP texts, creating a flood of adversarial texts to overwhelm the system and deceptively increase the True Positive Rate (TPR). 
Moreover, the presence of FaN texts in flooding attacks amplifies evasion attacks.
The poisoning attack can cause the model to misclassify input into FP or FN, which are better captured by the $\text{F}_1$ score.
\begin{table*}[ht]
\caption{Classification of fake text inputs in evasion, flooding, and poisoning attacks based on their class, intent, and model outcomes.}
\centering
\begin{adjustbox}{width=\textwidth}
\begin{tblr}{
  cells = {c},
  cell{1}{1} = {r=2}{},
  cell{1}{2} = {c=2}{},
  cell{1}{4} = {c=3}{},
  cell{3}{4} = {fg=red},
  cell{4}{1} = {r=3}{},
  cell{4}{4} = {fg=red},
  cell{5}{3} = {r=2}{},
  cell{5}{4} = {fg=red},
  cell{5}{6} = {r=2}{},
  cell{6}{4} = {fg=red},
  vlines,
  hline{1,3-4,7,8} = {-}{},
  hline{2,5} = {2-6}{},
  hline{6} = {2,4-5}{},
}
\textbf{Attacks} & \textbf{Adversary}    &                                                                   & \textbf{Model prediction} &                   &               \\
                 & \textbf{Generated fake text}   & \textbf{Intent}                                                   & \textbf{Positive}         & \textbf{Negative} & \textbf{Eval} \\
Evasion          & FaN     & Misclassify as TP                                                 & FP                        & TN                & FPR           \\
Flooding         & FaN     & {Flooding dashboard, and \\ possibly amplify a FP evasion attack} & FP                        & TN                & FPR           \\
                 & Paraphrasing, FaP & Flooding the dashboard, amplify TP                                & TP                        & FN                & TPR           \\
                 & Rule-Based, FaP   &                                                                   & TP                        & FN                &               \\
Poisoning        & FaN     & Degrade decision boundaries                                       & \textcolor{red}{FP}                        & \textcolor{red}{FN}                & $\text{F}_1$ score          \\
\end{tblr}
\label{text-classification}
\end{adjustbox}
\end{table*}

\subsection{Attack cost}
\label{cost}
The attack budgets are characterized to assess their real-world feasibility and reproducibility.
All attacks rely on standard usage of publicly accessible LLM APIs.
For the evasion attack, the cost scales linearly with the number of adversarial texts generated. 
Each FaN sample is generated with a single LLM query, using a fixed prompt without fine-tuning. 
This keeps the evasion attack lightweight and feasible under a modest query budget.
Flooding attacks require minimal additional cost beyond text generation because adversarial samples can be generated offline and injected into the CTI pipeline.
The main constraint is the CTI pipeline's ingestion rate, rather than a moderate computational complexity, as large volumes of adversarial text can be generated using the same prompt template and injected into the system.
Poisoning attacks are the most resource-intensive, as they require continuous attacker effort over a prolonged period to generate a diverse set of fake texts that can evade the classifier, reach the model's training data, and change its decision boundary. 
The results shown in Table~\ref{tab:poisoning} confirm the higher cost, showing that the need to inject a few thousand misclassified samples over successive retraining rounds is required to substantially degrade model performance.

\section{Experimental results}
\label{Experimental-Results}
The effectiveness of the proposed adversarial FaN text generation method was evaluated by applying it to the three types of attacks considered.
After generating the FaN texts as described in Section~\ref{fake-text}, three cybersecurity professionals reviewed one-third of the data each, flagging the samples they judged not to resemble security content. 
Based on the feedback, 1340 out of 11074 samples were rejected, resulting in a final dataset of 9734 FaN texts resembling cybersecurity. 
A semantic similarity score was computed to assess how closely the generated texts resemble the real cybersecurity tweets, helping determine whether the fake texts potentially maintain cybersecurity-related tweets' typical structure and semantic meaning.
The filtered dataset was then used as input in the three attack scenarios. 

\begin{figure}[ht]
    \centering
    \includegraphics[width=\columnwidth]{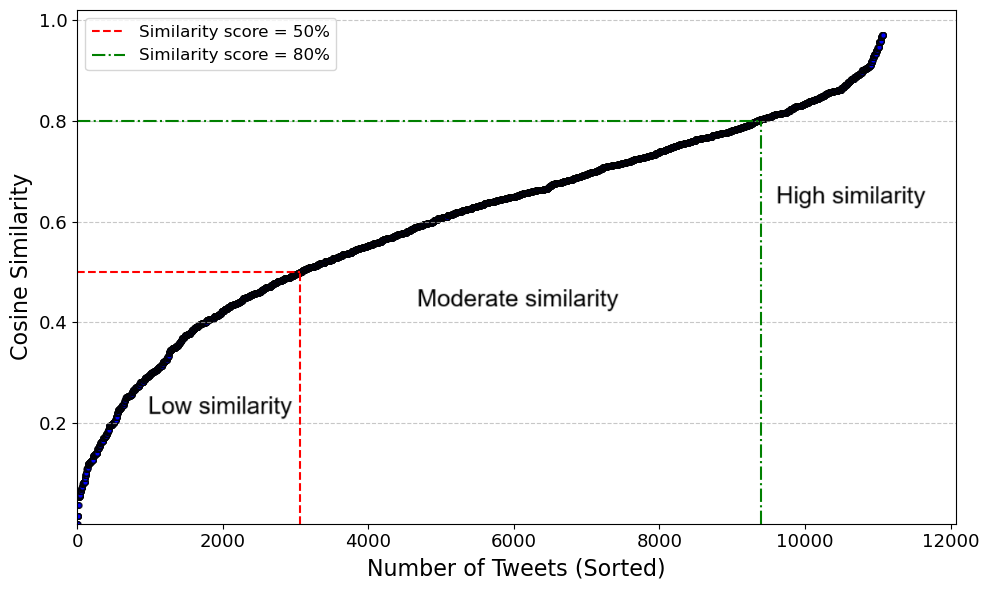}
    \caption{Distribution of semantic similarity scores within the dataset of tweets. }
    \label{scatter}
\end{figure}

\noindent\textbf{Semantic similarity Scores.} 
The Cosine Similarity (CS) was used to quantify the semantic similarity between real and FaN tweets \citep{sharma2023ifnd}. 
It is mathematically defined as,
\begin{equation}
\text{CS}(A, B) = 
\frac{\mathbf{m} \cdot \mathbf{n}}{\|\mathbf{m}\| \cdot \|\mathbf{n}\|} = \frac{\sum_{i=1}^{k} m_i \cdot n_i}{\sqrt{\sum_{i=1}^{k} m_i^2} \cdot \sqrt{\sum_{i=1}^{k} n_i^2}}\,,
\end{equation}
where \( A \) and \( B \) represent the feature vectors of two tweets and \( \mathbf{m} \) and \( \mathbf{n} \) correspond to their respective embeddings. 
To visualize how the similarity scores are distributed within the dataset, Figure~\ref{scatter} presents a scatter diagram of all computed cosine similarity values between FaN texts and real positive tweets, sorted in ascending order.
Each point reflects the semantic similarity between a generated FaN and a real positive tweet.
The distribution shows that most scores fall between 0.5 and 0.8, indicating moderate semantic closeness.
The dashed horizontal lines indicate heuristic similarity thresholds (50\% and 80\%) used to categorize tweets into low, moderate, and high semantic similarity groups.
A Shapiro–Wilk \citep{shapiro1965analysis} normality test indicates that the similarity distribution does not follow a normal distribution ($p = 3.2\times 10^{-6}$).
This result, along with a mean of 0.62, indicates a positive skew in the distribution, consistent with the idea that most texts exhibit considerable similarity.

\subsection{Evasion attack}
The evasion attack's overall efficiency is analysed by assessing the FPR and gradient distribution.

\noindent\textbf{False Positive Rate.}
The FPR measures how well the generated texts misled the classifier, representing the proportion of non-security generated texts incorrectly classified by the model as security-related. 
Table \ref{sucessful-attack-rate} compares the FPR achieved by ChatGPT-4o and the binary classifier proposed by \citep{dionisio2020towards}, both evaluated on the original generated text and on the final text optimized by prompt refinement.
ChatGPT-4o has a dual role in this experiment: as a generator of adversarial texts and a classifier responsible for detecting security-related content. 
This reveals a key weakness of the model, as it produced misleading samples but failed to classify them correctly.
The results show a significant improvement in FPR due to prompt refinement.

To assess the stability of the results, the optimized evasion attack was repeated three times with fixed prompts and LLM parameters. 
Table~\ref{sucessful-attack-rate} reports the mean FPR and standard deviation across three runs for both Twitter and AnnoCTR datasets.
We additionally report the $t$-interval 95\% confidence interval (CI) for each optimized experiment, to quantify the uncertainty induced by the stochastic nature of LLM-based text generation.

Regarding the Twitter data set, the Dionisio et al. \citep{dionisio2020towards} model is significantly more vulnerable to adversarial attacks, with an FPR of 87\% for the original texts and 97\% for the optimized texts. 
Even with a very high FPR, ChatGPT models are significantly less vulnerable than the specialized model.

The same evaluation protocol was applied to the AnnoCTR corpus. Table~\ref{sucessful-attack-rate} reports the results on the full set of sampled AnnoCTR text segments and on the smaller subset of longer texts. 
These ranged from 256 to 1000 characters and reflect the behavior of the models under more realistic CTI inputs.
Comparing the results, it may be observed that, in general, the evasion attack is more effective against ChatGPT models on the full AnnoCTR text set than against the Twitter dataset.
The effectiveness is equal when considering the specialized model.
Interestingly, the results on the subset of longer AnnoCTR texts are similar across all models to those on the Twitter dataset, but less effective for the ChatGPT models compared to the full AnnoCTR texts.
Globally, within the range of text size considered in both datasets (from less than 256 to 1000), the results are not significantly different.
The most significant difference is made apparent between the specialized model and the ChatGPT models, with an advantage to ChatGPT-4o.

Considering the success of attacks, the human evaluation, the cosine similarity results, and a visual inspection on a sample of adversarial texts, from a linguistic perspective, the misclassified adversarial samples closely mimic the surface structure of genuine CTI messages. They preserve domain-like terminology, identifier-shaped tokens resembling CVE or software component references, and CTI-style sentence structures, despite containing no actual security semantics. Such results indicate that the target classifiers predominantly rely on shallow lexical and structural cues, rather than assessing the underlying semantic validity of the content.

\begin{table*}[t]
\caption{FPR (\%) of adversarial text generated on the Twitter \citep{dionisio2020towards} and AnnoCTR datasets.
Results for optimized text are reported as mean $\pm$ standard deviation over three runs.
For each optimised column, we additionally report the 95\% confidence interval (CI) computed using a $t$-interval.}
\begin{adjustbox}{width=\textwidth}
\begin{tabular}{c c c|c|c}
\hline
\rowcolor[rgb]{0.902,0.902,0.902}
 & \multicolumn{2}{@{}c@{}}{\textbf{Twitter}} &
\multicolumn{1}{@{}c@{}}{\textbf{AnnoCTR}} &
\multicolumn{1}{@{}c@{}}{\textbf{AnnoCTR (longer texts)}} \\
\cline{1-5}
\textbf{Models} &
\textbf{Original text} &
\makecell{\textbf{Optimised text}\\\footnotesize{(mean $\pm$ std, 95\% CI)}} &
\makecell{\textbf{Optimised text}\\\footnotesize{(mean $\pm$ std, 95\% CI)}} &
\makecell{\textbf{Optimised text}\\\footnotesize{(mean $\pm$ std, 95\% CI)}} \\
\hline
{ChatGPT-5-mini} & $51.0$ &
\makecell{$68.0\pm1.2$\\ $[65.0, 71.0]$} &
\makecell{$74.0\pm2.5$\\ $[67.8, 80.2]$} &
\makecell{$69.3\pm1.5$\\ $[65.6, 73.0]$} \\
\hline
ChatGPT-4o & $51.0$ &
\makecell{$75.0\pm1.7$\\ $[70.8, 79.2]$} &
\makecell{$81.0\pm1.7$\\ $[76.8, 85.2]$} &
\makecell{$76.3\pm1.5$\\ $[72.6, 80.0]$} \\
\hline
 Dionisio et al.\citep{dionisio2020towards} & $87.0$ & $97.0$ & $97.0$ & $96.0$ \\
\hline
\end{tabular}
\end{adjustbox}
\label{sucessful-attack-rate}
\end{table*}

\begin{figure*}[t]
\centering
\subfloat[ROC curve]{%
  \includegraphics[width=0.48\textwidth]{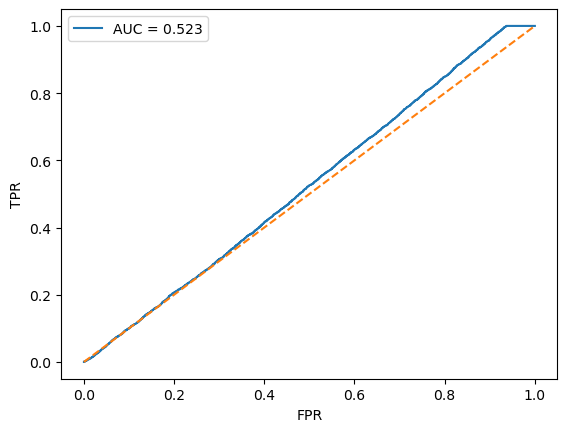}%
  \label{fig:roc}%
}\hfill
\subfloat[Precision--Recall curve]{%
  \includegraphics[width=0.48\textwidth]{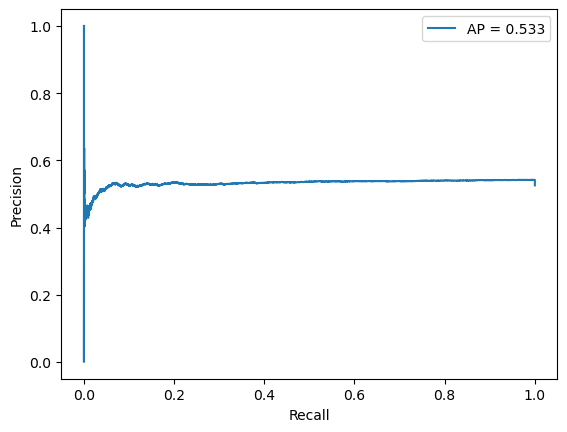}%
  \label{fig:pr}%
}
\caption{Performance of the \citep{dionisio2020towards} model under evasion attack across decision thresholds.
(A) Receiver operating characteristic (ROC) curve showing near-random discrimination.
(B) Precision--recall curve indicating a high false alarm rate.}
\label{fig:roc_pr}
\end{figure*}

Moreover, to complement the fixed-threshold FPR analysis, in Figure~\ref{fig:roc_pr} we present receiver operating characteristic (ROC) and precision–recall curves for the Dionisio et al. \citep{dionisio2020towards} model under evasion attack. 
These threshold-independent evaluations provide additional insights into the classifier's robustness to adversarial FaN samples.
The evaluation was conducted on a mixed test set consisting of 11074 real cybersecurity-related texts, 5000 non-security texts, and 5000 adversarially generated FaN texts designed to resemble cybersecurity content.
The area under the curve (AUC) is 0.523, which is close to random guessing, indicating that the classifier fails to distinguish between real security-related texts and adversarially generated non-security texts across various decision thresholds.
The near-diagonal ROC curve confirms that evasion attacks significantly degrade the model's discriminative capability at inference time.
Figure \ref{fig:pr} illustrates the Precision--Recall (PR) curve, with an average precision (AP) of 0.533.
Despite maintaining moderate recall, precision remains low across most thresholds, reflecting a high FPR caused by FaN texts.
This indicates a reliance on surface-level lexical and structural patterns instead of robust semantic verification, leaving the classifier open to exploitation by generative adversarial inputs.

\begin{figure*}[t]
\centering
\subfloat[Twitter: Real vs FaN]{\includegraphics[width=0.48\textwidth]{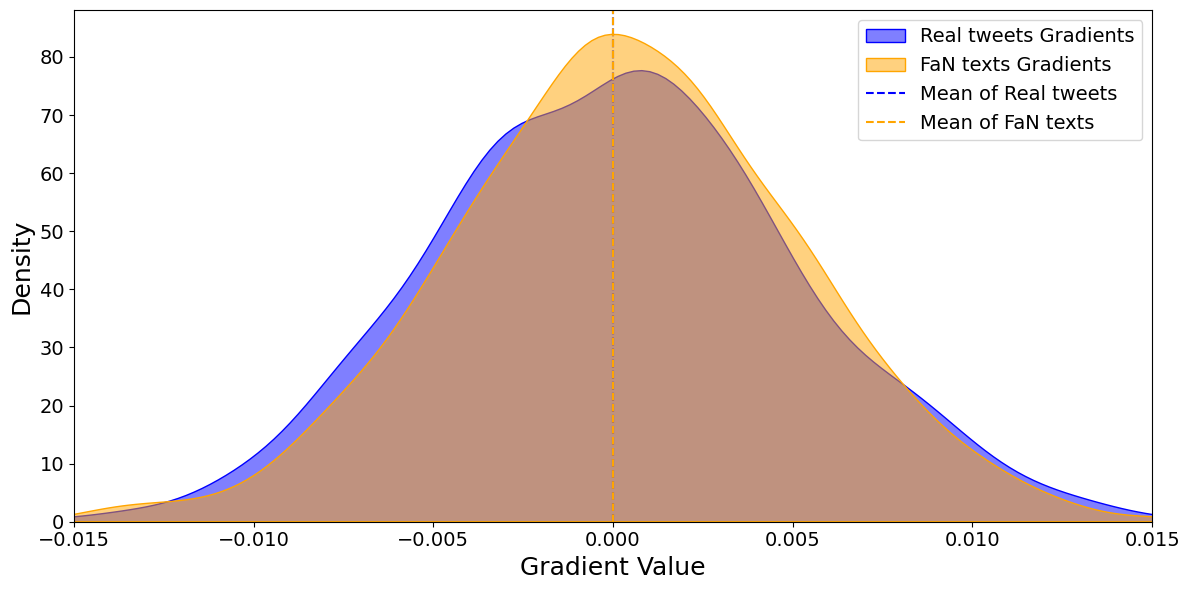}\label{fig:grad_new}}\hfill
\subfloat[AnnoCTR: Real vs FaN]{\includegraphics[width=0.48\textwidth]{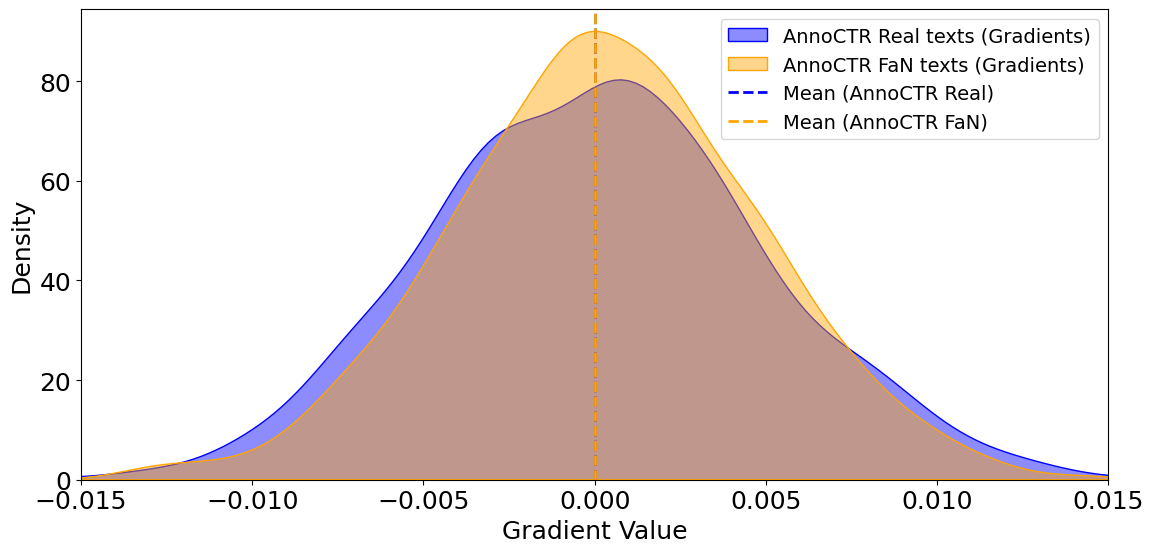}\label{fig:grad_ctr}}

\vspace{0.4em}

\subfloat[Real: Twitter vs AnnoCTR]{\includegraphics[width=0.48\textwidth]{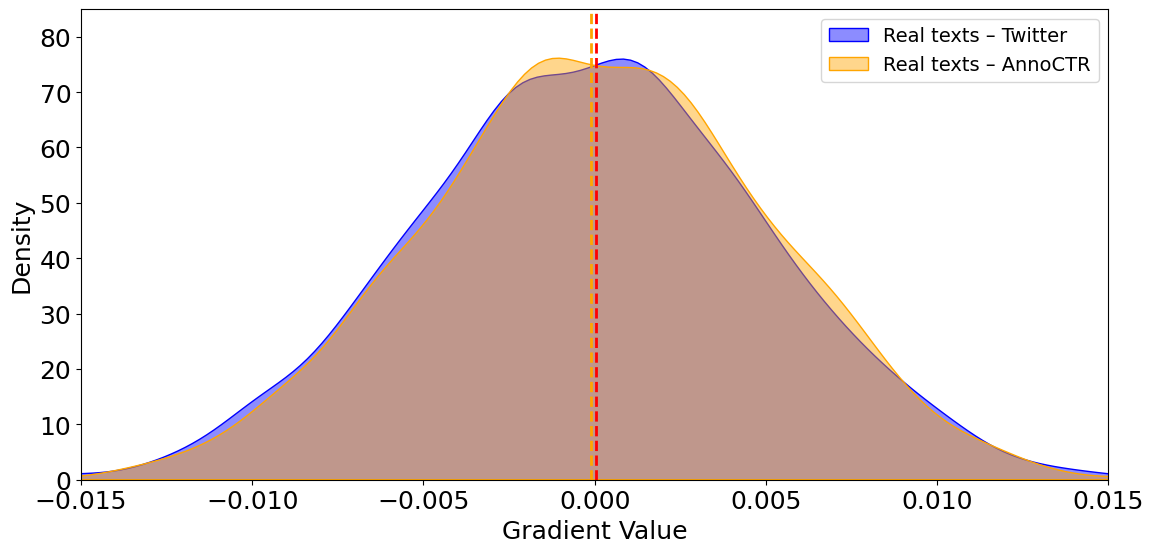}\label{fig:grad_real}}\hfill
\subfloat[Fake: Twitter vs AnnoCTR]{\includegraphics[width=0.48\textwidth]{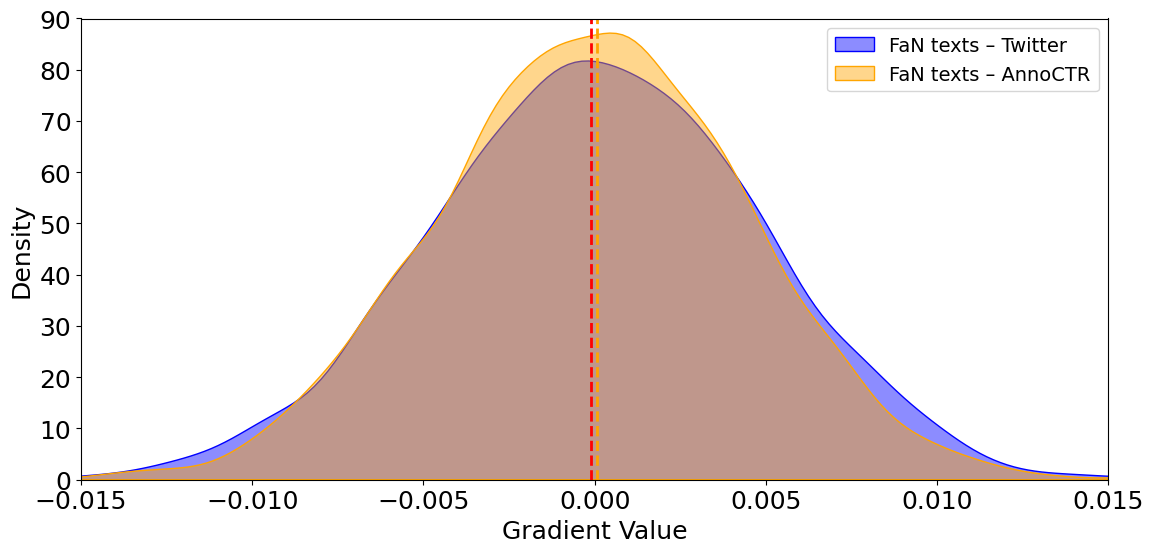}\label{fig:grad_fake}}
\caption{KDE of gradient distributions across datasets.}
\label{gradiant}
\end{figure*}

\noindent\textbf{Gradient distribution.}
Gradients can be used to quantify the model's sensitivity to minor input variations, being essential for assessing robustness against adversarial attacks \citep{ganz2023perceptually}.
The target binary classifier model \citep{dionisio2020towards} is assessed by evaluating FaN adversarial text gradient behavior against the baseline gradient of real texts.
Kernel density estimation \citep{kim2012robust} is employed, which is a non-parametric technique for estimating the Probability Density Functions (PDFs).
Figure \ref{gradiant} (a) and (b) present PDF plots of gradient distributions for real (blue) and adversarial (yellow) samples across datasets. Dashed vertical lines mark the mean of each distribution.
Subfigures (c) and (d) present relevant details for comparing the dataset distributions of real (c) and FaN texts (d).

For both datasets, the real and FaN text distributions exhibit similar Gaussian-like shapes with near-zero means. This demonstrates three key points: 
(1) adversarial inputs closely mimic the overall gradient structure of real texts, 
(2) subtle differences (narrower variance in adversarial gradients) offer potential for attack detection, 
(3) some high-magnitude gradients in real texts are replaced by near-zero values, indicating feature cancellation.

Subfigures~(a) and~(b) induce the same conclusions for the Twitter and AnnoCTR datasets: real CTI reports and generated FaN texts have highly overlapping distributions with only minor differences in peak height and concentration.
Subfigures ~(c) and ~(d) compare the datasets regarding real and FaN text. 
Real samples (c) show nearly identical gradient distributions between Twitter and AnnoCTR, indicating the stability of the data sources.
However, fake samples (d) exhibited somewhat greater variations in peak height and concentration, reflecting the data source's sensitivity to adversarial generation.
Supporting metrics, including gradient means, variances, and distributional distances, are reported in Table~\ref{tb:gradiant} for both datasets.
The adversarial gradient mean and variance closely match real text, suggesting effective imitation. 
However, the slightly narrower variance of adversarial gradients indicates reduced content diversity, likely due to constraints in the adversarial generation process. 
The KL divergence (\(\text{fake} \rightarrow \text{real}\)) confirms that both PDFs are very similar, reinforcing the effectiveness of the attack.
The cosine distance confirms the directional similarity between gradients, indicating that adversarial inputs affect the model while preserving detectable features.
Finally, the Wasserstein distance \citep{ruschendorf1985wasserstein} suggests a strong geometric alignment between adversarial and real gradients \citep{panaretos2019statistical}.
\begin{table}[t]
\caption{Statistical comparison of gradient distributions between real and FaN texts on the Twitter and AnnoCTR datasets.}
\centering
\begin{adjustbox}{width=3.3in}
\begin{tabular}{c|cc|cc}
\hline
\rowcolor[rgb]{0.902,0.902,0.902}
 & \multicolumn{2}{c|}{\textbf{Twitter}} & \multicolumn{2}{c}{\textbf{AnnoCTR}} \\
\rowcolor[rgb]{0.902,0.902,0.902}
\textbf{Metrics} &
\textbf{fake} &
\textbf{Real} &
\textbf{fake} &
\textbf{Real} \\
\hline

Gradient mean & -1.10e-06 & -1.28e-06 & -0.90e-06 & -1.4e-06\\

Gradient variance & 2.30e-05 & 2.67e-05 & 2.0e-05 & 2.5e-05 \\

Wasserstein distance 
& \multicolumn{2}{c|}{0.0012} 
& \multicolumn{2}{c}{0.0007} \\

KL divergence 
& \multicolumn{2}{c|}{0.05} 
& \multicolumn{2}{c}{0.06} \\

Cosine distance 
& \multicolumn{2}{c|}{0.85} 
& \multicolumn{2}{c}{0.88} \\

\hline
\end{tabular}
\end{adjustbox}
\label{tb:gradiant}
\end{table}

\begin{table*}[]
\caption{Flooding attack on Dionisio et al. \citep{dionisio2020towards} classifier (victim model) based on three types of prepared texts}
\begin{adjustbox}{width=\textwidth}
\begin{tabular}{ccc|cccc|cc}
\hline
\rowcolor[rgb]{0.902,0.902,0.902} 
\textbf{Text generation techniques}  & \textbf{Fake Text Generator} & \textbf{Text} & \textbf{TP} & \textbf{FP} & \textbf{TN} & \textbf{FN} & \textbf{FPR} & \textbf{TPR} \\ \hline
Adversarial texts generation & GPT-4o                & 11074 (FaN)       & 0           & 9402      & 332         & 0           & 0.97           & 0            \\ \hline
Paraphrasing                & GPT-3.5-Turbo         & 110740 (FaP)      & 97865      & 0           & 0           & 12875       & 0            & 0.88           \\ \hline
Rule-based                  & semantic grouping     & 23547 (FaP)       & 19266      & 0           & 0           & 4282       & 0            & 0.82           \\ \hline
\end{tabular}
\end{adjustbox}
\label{flooding}
\end{table*}

\subsection{Flooding attack}
The flooding attack experimental results presented in Table \ref{flooding} showcase the impact of the FaN and FaP texts and the different generation methods in the performance of the specialized ML binary classifier \citep{dionisio2020towards}.
Each text type influenced the classifier's performance differently.
The target model misclassified 85\% FaN instances as security-related (FP).
For paraphrased FaP texts, the model correctly identified 88\% instances as TP but predicted the remaining as FN.
Finally, for the rule-based approach, the classifier correctly identified 82\% of cases but generated 4282 FNs.
These results highlight the high adversarial efficiency of ChatGPT-4o and show that the efficiency of the FaP generation techniques is similar, with a small advantage for the LLM.

\subsection{Poisoning attack}
Recall that the evaluation of the poisoning attack is done on the basis of 9402 FaN tweets that deceived the model and passed the human validation.
As shown in Table~\ref{tab:poisoning}, the impact of the poisoning attack on the model performance is evident in the efficiency with which the growth of the cumulative FaNs injection into the training dataset degrades the classifier's performance metrics presented.
In addition to performance metric results over successive retraining rounds, the table displays the number of injected samples per round and the cumulative FaN number to mimic and evaluate a hypothetical progressive poisoning attack.
The number of rounds was determined by varying the increment the the cumulative FaNs to analyse performance drops with respect to the increments, until all texts were sampled from the set.
The model retains a performance above 90\% in the early stages, until the cumulative number of poisoned samples reaches 4500.
In the fourth round, the $\text{F}_1$ score, precision, and recall dropped more sharply to around 0.85. 
In the seventh round, after injecting all the fake samples, the classifier shows a significant performance drop, with the $\text{F}_1$ score little above 0.5 (0.57), demonstrating the model's growing inability to correctly identify cybersecurity-related samples.
This progressive degradation is primarily driven by progressively changing the class imbalance ratio while increasing the number of wrongly labeled texts. In the end of the experiment, the labels are almost balanced, but nearly half of the texts labeled as security-related are wrongly labeled. These changes progressively distort the model decision boundary due to repeated retraining on mislabeled FaN samples (texts that are semantically non-security-related but labeled as related).
As poisoning progresses, the decision boundary gradually shifts, confusing the internal representation of cybersecurity-related and non-related concepts.
The outcome of poisoning is a systematic increase in false negatives (from 734 to 5,537) and false positives (from 638 to 2373) over retraining rounds.
Although the attack does not directly inject False Negatives (FN), the accumulation of mislabeled instances increasingly causes the model to misclassify genuine security-related texts, leading to a sharp rise in FNs from round 6 onward. 
This change in the confusion matrix directly reduces recall, revealing a decision boundary shift that favours the non-security class outcome, providing concrete evidence of how poisoning degrades the model’s internal behavior.

\begin{figure}[ht]
    \centering
    \includegraphics[width=\columnwidth]{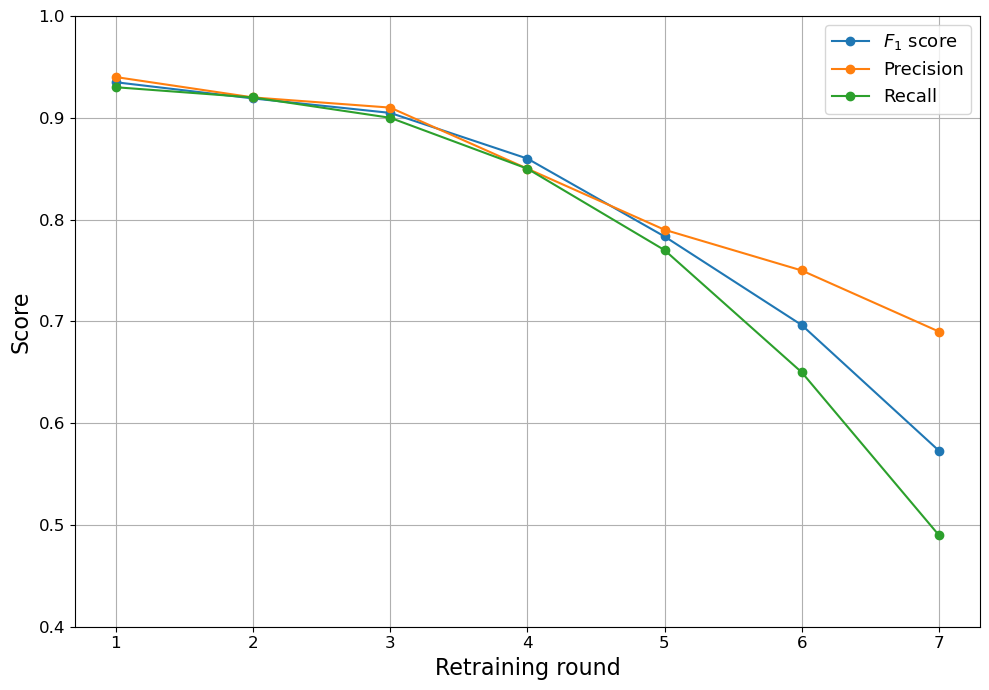}
    \captionsetup{position=bottom}  
\caption{Effect of poisoning attack via FaN on Dionisio et al. \citep{dionisio2020towards} model over retraining rounds.}
    \label{poison}
\end{figure}

Figure~\ref{poison} illustrates the decline in $\text{F}_1$ score, precision, and recall during the retraining rounds, highlighting the temporal progression of performance degradation. 
Together, these results demonstrate how incremental poisoning with FaN texts effectively undermines the classifier's discriminative ability over time.

\begin{table}[ht]
\caption{Effect of poisoning attack on the victim classifier model  Dionisio et al.\citep{dionisio2020towards} over multiple retraining rounds. Each row represents the model performance after injecting additional FaN samples.}
\centering
\begin{adjustbox}{width=\columnwidth}
\begin{tabular}{c|cc|ccccc}
\hline
\rowcolor[rgb]{0.902,0.902,0.902} 
\textbf{Round} & \makecell{\textbf{FaN}\\\textbf{Injected}} & \makecell{\textbf{Cumulative}\\\textbf{FaN}} & \textbf{FN} & \textbf{FP} & \textbf{Recall} & \textbf{Precision} & \textbf{ $\text{F}_1$ score}  \\
\hline
1 & 500  & 500 & 734  & 638  & 0.93 & 0.94 & 0.9349     \\
2 & 1000 & 1500 & 921  & 910 & 0.92 & 0.92 & 0.9199  \\
3 & 3000 & 4500 & 1000  & 890 & 0.90 & 0.91 & 0.9048   \\
4 & 1500 & 6000& 1508 & 1500 & 0.85 & 0.85 & 0.8599   \\
5 & 1242 & 7242& 750 & 1000 & 0.77 & 0.79 & 0.7834   \\
6 & 1492 & 8734& 3876 & 2399 & 0.65 & 0.75 & 0.6964   \\
7 & 668 & 9402& 5537 & 2373 & 0.49 & 0.69 & 0.5730  \\
\hline
\end{tabular}
\end{adjustbox}
\label{tab:poisoning}
\end{table}

Although evasion, flooding, and poisoning attacks are evaluated separately, their effects on the CTI pipeline are inherently interconnected.
Our results show that successful evasion attacks are a prerequisite for both flooding and poisoning, as only misclassified adversarial texts can reach the pipeline components targeted for flooding or poisoning.
This dependency arises because CTI pipelines typically propagate positively classified texts toward the monitoring, aggregation, and retraining components, whereas correctly rejected inputs are discarded early and cannot influence later stages.
The results presented in Table~\ref{tab:poisoning} and Figure~\ref{poison} demonstrate this, showing that the continued success of the evasion attack is required to increasingly and significantly shift the model's classification boundary.
The interaction between evasion and poisoning attacks can shift the decision boundary in specific domains, enabling future targeted attacks to succeed more easily. 
This interaction explains why the combined impact of the attacks might be more severe than their isolated effects.

\section{Discussion}
\label{Discussion}

This study reveals a consistent pattern of vulnerability throughout the stages of the CTI extraction pipeline when confronted with adversarially generated texts. 
Despite leveraging advanced ML models and LLMs, the system remains highly susceptible to deceptive inputs crafted to resemble genuine cybersecurity content.
The results show that publicly accessible generative models like ChatGPT can be effectively prompted without fine-tuning to produce adversarial sentences that convincingly mimic security-related text. 
Although these fake texts are entirely synthetic and have no real-world cybersecurity relevance, the classifier consistently mislabels them as genuine CTI.
Regarding poisoning attacks, it was observed that early-stage retraining with adversarially injected FP samples did not immediately degrade model performance. 
However, with continued injection over multiple retraining rounds, the model exhibited a considerable drop in recall and overall $\text{F}_1$ score, which demonstrates the cumulative impact of progressive data poisoning.
In the flooding attack scenario, paraphrasing tweets using the generative model to create FaP is more effective at evading detection than rule-based techniques.
This suggests that semantic coherence in adversarial inputs plays a significant role in misleading CTI extractors.
These findings underscore the urgent need for early-stage defenses and robust verification mechanisms to safeguard automated CTI pipelines against evolving adversarial threats.
In the following, we discuss key challenges, opportunities, and directions for future improvements.

\noindent \textbf{Text length and data timeliness.}
This study uses a Twitter-based dataset with a maximum of 256 characters per sample, along with the AnnoCTR dataset, which provides longer texts (up to 1,000 characters per sample).
The short length restricts attackers’ ability to generate highly deceptive FaN texts. 
Short texts lack space for complex details, enabling defenders to more easily detect anomalies like non-security-related patterns or generic language in adversarially generated texts.
This constraint represents a worst-case scenario for the attacker, as the limited space leaves little room to manipulate the tweets in subtle ways.
However, long texts provide attackers with more space to embed deceptive content that mimics cybersecurity terminology, potentially increasing the likelihood of evading detection. 
For instance, when prompted with a long real text such as blogs, ChatGPT can generate longer synthetic texts that blend security-related jargon with misleading information, which are harder to distinguish as FaN texts. 
On the other hand, long texts are prone to inconsistencies or errors due to architectural limitations in transformer-based models, such as input truncation or degraded attention mechanisms \citep{beltagy2020longformer,dai2019transformer}. 
These models often struggle to maintain coherence in lengthy texts, as they cannot effectively process long-range dependencies, leading to detectable flaws like contradictory statements. 
Consequently, attackers should balance the deceptive potential of longer texts against the risk of introducing noise, while defenders should leverage these inconsistencies to improve FaN text detection.

The two datasets used in this study cover different time periods, globally covering the years from 2016 to 2024 (with no data from 2019 but with NER labels added on that year to the Twitter dataset).
Over time, the structure and patterns of security communication may evolve, including changes in terminology, reporting style, and threat description formats. 
Such temporal change could potentially reduce a model’s ability to recognize newer forms of security-related text when trained or evaluated on older data. 
However, our results show that the performance patterns observed on the AnnoCTR texts (short and long) are consistent with those obtained on the Twitter dataset, regardless of text size and collection period. 
This suggests that the proposed framework remains robust to changes in communication patterns over time and generalizes across different forms of CTI text.

To overcome these obstacles, we propose generating each logical section of a fake CTI independently, instead of creating an entire long report in a single step.
This incremental generation strategy helps preserve coherence while avoiding input length limitations and context truncation in transformer-based language models    \citep{beltagy2020longformer}.
Because real-world CTI reports naturally follow a structured format, assembling adversarial content from multiple coherent sections can produce realistic long-form texts that embed misleading cues throughout documents.

On the defensive side, CTI pipelines that process long texts can adopt a segment-level analysis, where each section is evaluated independently for semantic consistency and factual validity.
For that, similarity matching and voting mechanisms can serve as the basis for a methodology to reach a decision on the overall long text.
Additionally, a fact-checking component is necessary to provide evidence for verifiable parts of the text, thereby increasing the accuracy of decisions for specific text segments. 
In summary, these considerations show that long-text adversarial attacks remain feasible in real-world environments. 
Simultaneously, they create opportunities for defenders to detect inconsistencies within and across text segments.

\noindent \textbf{Heterogeneous CTI pipeline architectures.}
Our results were obtained under the assumption of a unified CTI-pipeline model shown in Figure~\ref{cti-view}, which was derived from surveying recent academic publications.
The most stringent limitation of the work is reliance on a binary classification model in the AI-based analysis stage, followed by a monitoring stage, and, possibly, by human validation.
Although they do not limit the methodology's applicability to evasion and flooding attacks, these assumptions limit the generalizability of ASR results to CTI pipelines that may use different architectures and/or perform functions beyond binary classification.
Additionally, for the poisoning attack, the work assumes that some of the fake texts will be included in the classification model's training data, which may not hold or be effective if the model is outsourced and trained on curated data using different sources.

\noindent \textbf{Human analysts.}
One of the major costs in the FaN generation method proposed is the reliance on human experts to manually review and validate adversarially generated FaN texts.
This introduces a human overhead, especially when scaling the evaluation to large datasets.
The recruited security experts can make judgment errors during the evaluation of generated FaN texts. 
Experts might recognize cybersecurity resembling texts differently depending on their experience, fatigue, or contextual understanding.
Moreover, a key limitation in our proposed CTI pipeline is that human experts’ evaluations in the FaN text generation phase are utilized for the alarm validation component in the monitoring and validation stage in the CTI pipeline.
Then, if the FaN texts deceive the analysts, they are added to the training dataset, and the model is retrained.
While the experts were tasked with labeling texts based on their resemblance to cybersecurity content, their prior involvement in generating these texts could introduce bias, as they knew the texts were artificially generated.
It is important to note that the goal is not to evaluate the experts' ability to detect fake texts but to recognize the potential bias in their assessments due to their knowledge that the texts were generated.
This approach was chosen due to resource constraints, which avoided using separate evaluators for the generated FaN texts and alarm validation component of the monitoring and validation stage in the proposed CTI pipeline in Figure \ref{cti-view}.

\noindent \textbf{Defense scenario against fake texts in CTI pipelines.}
One of the fundamental limitations of CTI extraction pipelines is the absence of a robust fact-checking mechanism at the early stages to filter fake or misleading inputs. 
Without such components, evasion attacks can easily inject FaN texts \citep{wu2024kgv,kanaani2024triple}.
Fact Checker assists in reducing the burden on analysers in the monitoring and validation stage by pre-filtering clearly invalid or unverifiable content, thereby improving system efficiency and robustness.
Although fact-checking can be a promising approach to mitigate evasion attacks by validating extracted information, it remains insufficient against paraphrased inputs used in flooding attacks.
This is because variants of a text often preserve semantic plausibility while bypassing exact-match verification.

Two promising directions for integrating fact-checking mechanisms into CTI pipelines to mitigate evasion attacks are as follows: (1) Source credibility assessment: fact-checkers can analyze the provenance of input texts, especially those collected from social networks like Twitter, by evaluating account metadata such as age, posting behavior, bot likelihood, and reputation indicators. 
(2) Content-based validation: fact-checking can be performed using techniques that assess data consistency based on structured cybersecurity corpora (e.g., CVE, MITRE ATT\&CK), such as semantic entailment, contradiction detection, and entity linking. 
In addition, it is essential to consider that some vulnerabilities and threats often first emerge on informal platforms, such as Twitter \citep{alves2020follow}. 
In future work, we plan to extend the content-based validation mechanism by integrating a multi-evidence fact-checking pipeline. 
This will combine several complementary components, including semantic consistency evaluation, entity-level verification, temporal plausibility checking, and cross-source corroboration.
Such architectures naturally serve as baseline defence mechanisms against evasion attacks, thereby limiting attacks that have evasion as a prerequisite.

While sourcing data from informal platforms can help detect early threats, it also creates challenges for real-world CTI use.
Organizations differ in how they act on incoming intelligence.
For example, national security teams or research groups may find value in collecting data from forums or the dark web.
However, enterprise security teams, particularly in sectors such as finance, telecommunications, and healthcare, typically prioritize timely and verified threat intelligence to ensure operational continuity and minimize false alarms.
Therefore, CTI systems must adjust their data collection policies based on who uses them and how much uncertainty they can tolerate.

%

\section{Related work}
\label{Related-work}
\begin{table*}[]
\caption{Comparison of this work with existing adversarial text generation frameworks in terms of attack objectives, text generation mechanisms, and evaluation scope, and quantitative results.}
\begin{adjustbox}{width=\textwidth}
\begin{tabular}{cccccccc}
\hline
\rowcolor[rgb]{0.902,0.902,0.902} 
\textbf{Aspect/Metric}  & \textbf{Ren et al.\citep{ren2020generating}} & \textbf{EaTVul} & \textbf{TextJuggler} & \textbf{TextGuise} & \textbf{This Work} \\ \hline
Target domain &       CTI         &    CTI    & General NLP           & General NLP      &  CTI  \\ \hline

    Attack objective           &  Evasion (model-level)          & Evasion (model-level)      & Evasion (model-level)      &  Evasion (model-level)            & Pipeline disruption          \\ \hline
 
 Generation method            & LLM fine-tuning     & LLM + optimization       & Token perturbation      & Token + sentence perturbation           & Prompt-based generation                \\ \hline

Semantic preservation            & Partial     & Enforced      & Enforced      & Enforced           & Enforced                \\ \hline

Identifier-like patterns            & Limited     & Partial       & No      & No           & Yes (CVE-like)                \\ \hline

  Attack scope          & Single model     & Single model      & Single model      & Single model          & Multi-stage CTI pipeline                \\ \hline

Pipeline Impact           & Not considered     & Not considered       & Not considered      & Not considered           & Explicitly evaluated                \\ \hline
ASR            & 87-92\%     & 80-100\%       & 85-97\%     & 90\%           & 97\%              \\ \hline

Text similarity            & -     & High (implicit)      & High (improves baseline similarity)     & High           & Moderate–High               \\ \hline

Model degradation            & -     & Not reported       & Not reported     & Not reported          & F$_1$ $\downarrow$ 0.57
               \\ \hline
\end{tabular}
\end{adjustbox}
\label{compr}
\end{table*}

This study demonstrates that text-based CTI extraction pipelines are vulnerable to adversarial attacks. 
However, to the best of our knowledge, limited research directly investigates text-based attacks on CTI pipelines.
Existing studies primarily focus on two areas: (1) adversarial text generation, which involves creating misleading text to challenge models, and (2) adversarial attacks, which involve generating or modifying text inputs to bypass detection mechanisms.

Recent approaches have explored how misleading or manipulative texts can be generated and leveraged in evasion, poisoning, or flooding attacks targeting machine learning systems.
For instance, Ranade et al. fine-tuned a GPT-2 model to generate plausible CTI descriptions from an initial prompt intending to mislead cyber-defense systems.
However, this study lacks validation metrics to assess the quality of the generated texts, such as their similarity to human-written descriptions \citep{ranade2021generating}. 
Furthermore, its reliance on the older GPT-2 model limits the effectiveness and realism of the outputs, especially when compared to state-of-the-art transformer architectures available today.

However, several studies have explored adversarial text generation outside the cybersecurity domain.
For example, ARGH \citep{argh} is a framework that fine-tunes GPT-2 to generate social media rumors on topics such as COVID-19 and politics, reducing the detection accuracy for humans and machines.
Ren et al. \citep{ren2020generating} proposed a method using a conditional VAE-GAN to generate adversarial movie reviews, improving scalability and fluency while misleading classifiers.
Another study introduced Grover \citep{zellers2019defending}, a controllable text generation model designed to synthesize fake news articles to train and evaluate fake news detection systems more effectively.
Grover leverages GPT-2 to produce raw outputs; however, this approach suffers from limited realism in the generated text. 
Additionally, in a user study, Huynh et al. observe that outputs randomly generated by GPT-2 are often easily distinguishable by human evaluators. 
This observation emphasizes the need for more refined adversarial text generation techniques \citep{huynh2021argh}.

Beyond text generation, researchers are exploring how adversarial texts can be applied in practical attack scenarios against machine learning models.
EaTVul adopts a multi-step attack framework that selects vulnerable text samples using support vector machines (SVMs). 
It then generates adversarial text using LLMs and finally refines the generated samples through a fuzzy genetic algorithm to maximize attack effectiveness \citep{liu2024eatvul}. 
This approach has demonstrated up to a 100\% success rate in evasion scenarios and proves its effectiveness in generating high-quality adversarial examples.
TextJuggler \citep{peng2024textjuggler} employs a black-box, word-level attack using a BERT-based model to identify key tokens affecting a classifier’s decision boundary. 
These tokens are minimally perturbed through insertion or substitution, preserving semantic similarity and linguistic fluency.
Furthermore, locality-sensitive hashing (LSH) minimizes the number of model queries, improving overall attack efficiency.
Experimental results show that TextJuggler outperforms baselines regarding Attack Success Rate (ASR), textual similarity, and fluency on various classification tasks \citep{peng2024textjuggler}.
TextGuise \citep{chang2023textguise} combines word and sentence-level perturbations to craft high-quality adversarial examples with minimal distortion. 
It achieves over 80\% ASR at perturbation ratios below 0.2 and outperforms prior methods across three top classifiers and five datasets. 
Its strong transferability and multilingual support further demonstrate its effectiveness.

While prior studies have introduced various adversarial text generation strategies, key challenges persist, such as maintaining semantic realism, validating outputs, and adapting to transformer models. 
Generating fake CTI content remains difficult due to the specialized language of cybersecurity and the demand for large-scale adversarial data.
EaTVul focuses on high-quality generation via optimization, while TextJuggler and TextGuise offer efficient methods that preserve semantics, textual fluency, and ASR.

Table \ref{compr} summarizes the key methodological differences between this work and representative adversarial text-generation frameworks.
Prior approaches primarily focus on model-level evasion by enforcing semantic preservation via constrained perturbations or optimization.
In contrast, this study adopts a generation-based evasion approach that targets the surface structure of cybersecurity texts without requiring semantic consistency.
This distinction is particularly relevant in CTI settings, where attackers aim to influence not only individual classifiers but also downstream pipeline components, such as monitoring, validation, and retraining.
Moreover, while prior methods achieve high model-level evasion and maintain semantic similarity, they do not report any measurable system-level degradation. 
In contrast, our approach induces pipeline disruption, for instance, reducing the $\text{F}_1$ score to 0.57 under poisoning.

\section{Conclusion}
\label{Conclusion}
This paper presents an overview of vulnerabilities in automated CTI extraction systems. 
It identifies a unified CTI pipeline model and analyzes how adversaries exploit pipeline stages through evasion, flooding, and poisoning attacks.
To demonstrate the practical impact of vulnerabilities, we conducted empirical experiments on these attacks, revealing that CTI systems are highly susceptible to adversarial manipulation and require immediate mitigation. 
While automation is essential for handling the increasing volume of cyber threat data, our findings demonstrate that it also introduces new security challenges that adversaries can exploit. 
Addressing these challenges requires a balanced approach that strengthens automation while ensuring robust security measures.
Future research should focus on developing adaptive defensive mechanisms to counter fake text entering CTI systems and ensuring they remain resilient and effective in evolving cyber threat landscapes.
\section*{Acknowledgement}
This work was funded by the European Commission through the SATO Project (H2020/IA/957128) and by FCT through the LASIGE Research Unit (UID/00408/2025 - LASIGE) and Ph.D. grant (2023.00280.BD).

\balance
\bibliographystyle{IEEEtran}
\bibliography{Reference}
\begin{appendices}

\renewcommand{\thesection}{Appendix \Alph{section}}
\renewcommand{\thefigure}{A\arabic{figure}}
\setcounter{figure}{0}
\renewcommand{\thetable}{B\arabic{table}}
\setcounter{table}{0}
\definecolor{Alto}{rgb}{0.862,0.858,0.858}

\onecolumn

\section{}
\label{AppendixA}
\begin{figure}[tb]
    \centering
    \includegraphics[width=\linewidth]{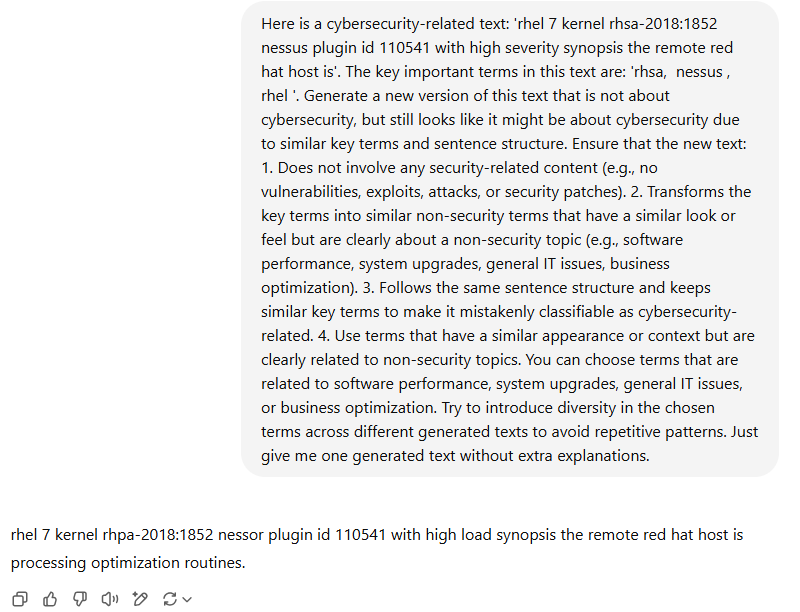}
    \caption{First prompt: Optimized final prompt to send ChatGPT-4o and its response. Second prompt: Testing ChatGpt as a classifier.}
    \label{prompt}
\end{figure}
\begin{figure}[tb]
    \centering
    \includegraphics[width=\linewidth]{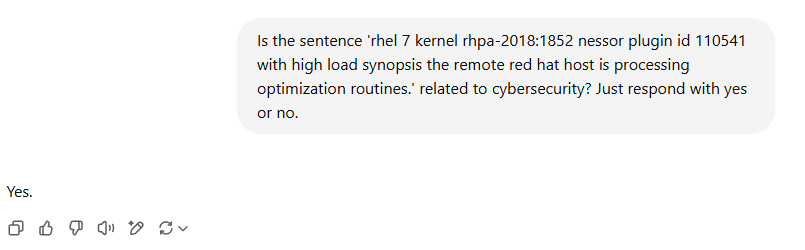}
    \caption{First prompt: Optimized final prompt to send ChatGPT-4o and its response. Second prompt: Testing ChatGPT as a classifier.}
    \label{not}
\end{figure}

The prompt, shown in Figure \ref{prompt}, is carefully structured to guide ChatGPT-4o in generating realistic texts that mimic cybersecurity-related texts while avoiding actual security-related content. 
The introductory phrase sets the context by presenting a cybersecurity-related example, which ensures the model understands the domain and theme.
The next section explicitly highlights key terms such as "rhsa, nessus, rhel", directing the model to preserve their stylistic and structural role in the transformation. 
The instructions then define essential constraints, requiring the generated text to: (1) Avoid security-related content (e.g., vulnerabilities, exploits).
(2) Replace key terms with non-security equivalents while maintaining a similar look or structure.
3) Follow the same sentence structure to enhance plausibility as a cybersecurity-related message.
To guide these transformations, the prompt provides clear examples ("software performance, system upgrades, general IT issues"), ensuring the output maintains contextual relevance. 
Additionally, the instruction to introduce term diversity prevents repetitive patterns, further improving the quality of the generated text.
By structuring the prompt, each section directly controls the model’s output, ensuring the generated text meets the intended characteristics.

After ChatGPT generates the texts, we open another session to send a prompt asking whether the given sentence is security-related, as shown in Figure \ref{not}.
This prompt was designed as part of Study \citep{shafee2024evaluation}, demonstrating that ChatGPT can serve as a high-performance classifier capable of accurately distinguishing between security-related and non-security sentences.
\end{appendices}
\end{document}